\documentclass[a4paper,manyauthors,nocleardouble,COMPASS]{cernphprep}

\RequirePackage[T1]{fontenc}

\RequirePackage{graphicx}
\RequirePackage{mathptmx}      % use Times fonts if available on your TeX system
\RequirePackage{flushend}
\RequirePackage[numbers,sort&compress]{natbib}
\RequirePackage[colorlinks,citecolor=blue,urlcolor=blue,linkcolor=blue]{hyperref}
\usepackage{tikz}
\usepackage{lineno}
\usepackage[normalem]{ulem} % added by Anatolii to easily cross out text
\usepackage{soul} % added by Anatolii to easy highlight with color
\usepackage{amsmath}
\usepackage{balance}
\usepackage{lastpage}

\renewcommand\vec{\mathbf}
\begin{document}
%\linenumbers
\begin{titlepage}
\PHnumber{2020--xxx}
\PHdate{\today}
\title{Spin Density Matrix Elements in Exclusive $\omega$ Meson
Muoproduction $^*$}

\Collaboration{The COMPASS Collaboration}                                                 
\ShortAuthor{The COMPASS Collaboration} 

%
%\author{
%{Draft version 2.0}
%}
%
%
%\institute{
% %First Address, Street, City, Country\label{addr1}
%}
%\date\today
%
%\maketitle
%
\begin{abstract}
We report on a measurement of 
Spin Density Matrix Elements (SDMEs) in hard  
exclusive
 $\omega$ meson muoproduction on the proton at COMPASS using 160~GeV/$c$ polarised $ \mu ^{+}$ and $ \mu ^{-}$ beams impinging on a liquid
 hydrogen target. The measurement covers the range 5.0~GeV/$c^2$
 $< W <$ 17.0~GeV/$c^2$, with the average kinematics
$\langle Q^{2} \rangle=$ 2.1~(GeV/$c$)$^2$, $\langle W \rangle= 7.6$~GeV/$c^2$,
and $\langle p^{2}_{\rm T} \rangle = 0.16$~(GeV/$c$)$^2$.
Here, $Q^2$ denotes the virtuality of the exchanged photon, $W$ the mass of the final hadronic system and $p_T$ the
transverse momentum of the $\omega$ meson  with
respect to the virtual-photon direction. 
The measured non-zero SDMEs for
the transitions of transversely polarised virtual photons to longitudinally polarised
vector mesons ($\gamma^*_T \to V_L$) indicate a violation of $s$-channel helicity conservation. Additionally, we observe a sizeable contribution of unnatural-parity-exchange (UPE) transitions 
that
decreases with increasing $W$. The  results provide
important input for modelling Generalised Parton Distributions (GPDs). In particular,
 they may allow to evaluate in a model-dependent way the contribution of
UPE transitions and assess the role of
parton helicity-flip GPDs 
in exclusive $\omega$ production.
\end{abstract}

\vfill
\Submitted{(to be submitted to Eur. Phys. J. C)}
\end{titlepage}
{
\pagestyle{empty}
%%%%%%%%%%%%%%%%%%%%%%%%%%%%%%%%%%%%%%%%%%%%%%%%%%%%%%%%%%%%%%%
%
% 2020_auththorlist.tex (updated 19.04.2020)
%
%%%%%%%%%%%%%%%%%%%%%%%%%%%%%%%%%%%%%%%%%%%%%%%%%%%%%%%%%%%%%%%
\section*{The COMPASS Collaboration}
\label{app:collab}
\renewcommand\labelenumi{\textsuperscript{\theenumi}~}
\renewcommand\theenumi{\arabic{enumi}}
\begin{flushleft}
%J.~Agarwala\Irefn{triest_i}\Aref{JA},
G.D.~Alexeev\Irefn{dubna}, %1
M.G.~Alexeev\Irefnn{turin_u}{turin_i},
A.~Amoroso\Irefnn{turin_u}{turin_i},
V.~Andrieux\Irefnn{cern}{illinois},
%N.V.~Anfimov\Irefn{dubna}, %2
V.~Anosov\Irefn{dubna}, %3
A.~Antoshkin\Irefn{dubna},
K.~Augsten\Irefnn{dubna}{praguectu}, %2 phd
W.~Augustyniak\Irefn{warsaw},
C.D.R.~Azevedo\Irefn{aveiro},
B.~Bade{\l}ek\Irefn{warsawu},
F.~Balestra\Irefnn{turin_u}{turin_i},
M.~Ball\Irefn{bonniskp},
J.~Barth\Irefn{bonniskp},
R.~Beck\Irefn{bonniskp},
Y.~Bedfer\Irefn{saclay},
J.~Berenguer~Antequera\Irefnn{turin_u}{turin_i},
J.~Bernhard\Irefnn{mainz}{cern},
M.~Bodlak\Irefn{praguecu},
%P.~Bordalo\Irefn{lisbon}\Aref{A},
F.~Bradamante\Irefn{triest_i},
A.~Bressan\Irefnn{triest_u}{triest_i},
%M.~B\"uchele\Irefn{freiburg},
V.E.~Burtsev\Irefn{tomsk},
W.-C.~Chang\Irefn{taipei},
C.~Chatterjee\Irefnn{triest_u}{triest_i},
M.~Chiosso\Irefnn{turin_u}{turin_i},
A.G.~Chumakov\Irefn{tomsk},
S.-U.~Chung\Irefn{munichtu}\Aref{B}\Aref{B1}
A.~Cicuttin\Irefn{triest_i}\Aref{C}
P.M.M.~Correia\Irefn{aveiro},
M.L.~Crespo\Irefn{triest_i}\Aref{C}
% ...
\end{flushleft}
%%%%%%%%%%%%%%%%%%%%%%%%%%%%%%%%%%%%%%%%%%%%%%%%%%%%%%%%%%%%%%%%%%%%%%%%%%%%%%%%%%%%%%%%%%%%%%%%%%%%%%%%%%%%%%%%%%%%%%%
%
% institutes
%
%%%%%%%%%%%%%%%%%%%%%%%%%%%%%%%%%%%%%%%%%%%%%%%%%%%%%%%%%%%%%%%%%%%%%%%%%%%%%%%%%%%%%%%%%%%%%%%%%%%%%%%%%%%%%%%%%%%%%%%
%\item \Idef{bielefeld}{Universit\"at Bielefeld, Fakult\"at f\"ur Physik, 33501 Bielefeld, Germany\Arefs{l}}
%\item \Idef{munichlmu}{Ludwig-Maximilians-Universit\"at M\"unchen, Department f\"ur Physik, 80799 Munich, Germany\Arefs{l}\Arefs{r}}
\begin{Authlist}
\item \Idef{aveiro}{University of Aveiro, I3N, Dept.\ of Physics, 3810-193 Aveiro, Portugal}
\item \Idef{bochum}{Universit\"at Bochum, Institut f\"ur Experimentalphysik, 44780 Bochum, Germany\Arefs{Q}\Aref{R}}
\item \Idef{bonniskp}{Universit\"at Bonn, Helmholtz-Institut f\"ur  Strahlen- und Kernphysik, 53115 Bonn, Germany\Arefs{Q}}
\item \Idef{bonnpi}{Universit\"at Bonn, Physikalisches Institut, 53115 Bonn, Germany\Arefs{Q}}
\item \Idef{brno}{Institute of Scientific Instruments of the CAS, 61264 Brno, Czech Republic\Arefs{S}}
\item \Idef{calcutta}{Matrivani Institute of Experimental Research \& Education, Calcutta-700 030, India\Arefs{T}}
\item \Idef{dubna}{Joint Institute for Nuclear Research, 141980 Dubna, Moscow region, Russia\Arefs{T1}}
\item \Idef{freiburg}{Universit\"at Freiburg, Physikalisches Institut, 79104 Freiburg, Germany\Arefs{Q}\Aref{R}}
\item \Idef{cern}{CERN, 1211 Geneva 23, Switzerland}
\item \Idef{liberec}{Technical University in Liberec, 46117 Liberec, Czech Republic\Arefs{S}}
\item \Idef{lisbon}{LIP, 1649-003 Lisbon, Portugal\Arefs{U}}
\item \Idef{mainz}{Universit\"at Mainz, Institut f\"ur Kernphysik, 55099 Mainz, Germany\Arefs{Q}}
\item \Idef{miyazaki}{University of Miyazaki, Miyazaki 889-2192, Japan\Arefs{V}}
\item \Idef{moscowlpi}{Lebedev Physical Institute, 119991 Moscow, Russia}
\item \Idef{munichtu}{Technische Universit\"at M\"unchen, Physik Dept., 85748 Garching, Germany\Arefs{Q}\Aref{D}}
\item \Idef{nagoya}{Nagoya University, 464 Nagoya, Japan\Arefs{V}}
\item \Idef{praguecu}{Charles University, Faculty of Mathematics and Physics, 18000 Prague, Czech Republic\Arefs{S}}
\item \Idef{praguectu}{Czech Technical University in Prague, 16636 Prague, Czech Republic\Arefs{S}}
\item \Idef{protvino}{State Scientific Center Institute for High Energy Physics of National Research Center `Kurchatov Institute', 142281 Protvino, Russia}
\item \Idef{saclay}{IRFU, CEA, Universit\'e Paris-Saclay, 91191 Gif-sur-Yvette, France\Arefs{R}}
\item \Idef{taipei}{Academia Sinica, Institute of Physics, Taipei 11529, Taiwan\Arefs{W}}
\item \Idef{telaviv}{Tel Aviv University, School of Physics and Astronomy, 69978 Tel Aviv, Israel\Arefs{X}}
\item \Idef{tomsk}{Tomsk Polytechnic University, 634050 Tomsk, Russia\Arefs{Y}}
\item \Idef{triest_u}{University of Trieste, Dept.\ of Physics, 34127 Trieste, Italy}
\item \Idef{triest_i}{Trieste Section of INFN, 34127 Trieste, Italy}
\item \Idef{turin_u}{University of Turin, Dept.\ of Physics, 10125 Turin, Italy}
\item \Idef{turin_i}{Torino Section of INFN, 10125 Turin, Italy}
\item \Idef{illinois}{University of Illinois at Urbana-Champaign, Dept.\ of Physics, Urbana, IL 61801-3080, USA\Arefs{Z}}
\item \Idef{warsaw}{National Centre for Nuclear Research, 02-093 Warsaw, Poland\Arefs{a} }
\item \Idef{warsawu}{University of Warsaw, Faculty of Physics, 02-093 Warsaw, Poland\Arefs{a} }
\item \Idef{warsawtu}{Warsaw University of Technology, Institute of Radioelectronics, 00-665 Warsaw, Poland\Arefs{a} }
\item \Idef{yamagata}{Yamagata University, Yamagata 992-8510, Japan\Arefs{V} }
%\item \Idef{retired}{Retired}
\end{Authlist}
%%%%%%%%%%%%%%%%%%%%%%%%%%%%%%%%%%%%%%%%%%%%%%%%%%%%%%%%%%%%%%%%%%%%%%%%%%%%%%%%%%%%%%%%%%%%%%%%%%%%%%%%%%%%%%%%%%%%%%%
%
% Notes
%
%%%%%%%%%%%%%%%%%%%%%%%%%%%%%%%%%%%%%%%%%%%%%%%%%%%%%%%%%%%%%%%%%%%%%%%%%%%%%%%%%%%%%%%%%%%%%%%%%%%%%%%%%%%%%%%%%%%%%%%
%\vspace*{-\baselineskip}
\renewcommand\theenumi{\alph{enumi}}
\begin{Authlist}
\item [{\makebox[2mm][l]{\textsuperscript{*}}}] Dedicated to the memory of Bohdan Marianski
\item [{\makebox[2mm][l]{\textsuperscript{\#}}}] Corresponding authors\\
{\it E-mail addresses}: Oleg.Denisov@cern.ch, Andrzej.Sandacz@ncbj.gov.pl, Fulvio.Tessarotto@cern.ch
%\item [{\makebox[2mm][l]{\textsuperscript{*}}}] Deceased
%\item \Adef{JA}{Present address: University of 27100 Pavia, Pavia, Italy}
%\item \Adef{A}{Also at Instituto Superior T\'ecnico, Universidade de Lisboa, Lisbon, Portugal}
\item \Adef{B}{Also at Dept.\ of Physics, Pusan National University, Busan 609-735, Republic of Korea}
\item \Adef{B1}{Also at Physics Dept., Brookhaven National Laboratory, Upton, NY 11973, USA}
\item \Adef{C}{Also at Abdus Salam ICTP, 34151 Trieste, Italy}
\item \Adef{D}{Supported by the DFG cluster of excellence `Origin and Structure of the Universe' (www.universe-cluster.de) (Germany)}
%\fntext[E]{Supported by CERN-RFBR Grant 12-02-91500}
\item \Adef{F}{Supported by the DFG Research Training Group Programmes 1102 and 2044 (Germany)}
\item \Adef{G}{Also at Chubu University, Kasugai, Aichi 487-8501, Japan}
\item \Adef{H}{Also at Dept.\ of Physics, National Central University, 300 Jhongda Road, Jhongli 32001, Taiwan}
\item \Adef{I}{Also at KEK, 1-1 Oho, Tsukuba, Ibaraki 305-0801, Japan}
\item \Adef{J}{Present address: Universit\"at Bonn, Physikalisches Institut, 53115 Bonn, Germany}
\item \Adef{K}{Also at Moscow Institute of Physics and Technology, Moscow Region, 141700, Russia}
\item \Adef{K1}{Deceased}
\item \Adef{L}{Also at Yerevan Physics Institute, Alikhanian Br. Street, Yerevan, Armenia, 0036}
\item \Adef{M}{Also at Dept.\ of Physics, National Kaohsiung Normal University, Kaohsiung County 824, Taiwan}
\item \Adef{M1}{Supported by ANR, France with P2IO LabEx (ANR-10-LBX-0038) in the framework ``Investissements d'Avenir'' (ANR-11-IDEX-003-01)}
\item \Adef{N}{Also at Institut f\"ur Theoretische Physik, Universit\"at T\"ubingen, 72076 T\"ubingen, Germany}
\item \Adef{N1}{Retired}
\item \Adef{N2}{Present address: Brookhaven National Laboratory, Brookhaven, USA}
\item \Adef{O}{Also at University of Eastern Piedmont, 15100 Alessandria, Italy}

\item \Adef{P}{Present address: RWTH Aachen University, III.\ Physikalisches Institut, 52056 Aachen, Germany}
\item \Adef{P1}{Present address: Universit\"at Hamburg, 20146 Hamburg, Germany}
\item \Adef{Q}{Supported by BMBF - Bundesministerium f\"ur Bildung und Forschung (Germany)}
\item \Adef{R}{Supported by FP7, HadronPhysics3, Grant 283286 (European Union)}
\item \Adef{S}{Supported by MEYS, Grant LM20150581 (Czech Republic)}
\item \Adef{T}{Supported by B.~Sen fund (India)}
\item \Adef{T1}{Supported by CERN-RFBR Grant 12-02-91500}
\item \Adef{U}{Supported by FCT, Grants CERN/FIS-PAR/0007/2017 and  CERN/FIS-PAR/0022/2019 (Portugal)}
\item \Adef{V}{Supported by MEXT and JSPS, Grants 18002006, 20540299, 18540281 and 26247032, the Daiko and Yamada Foundations (Japan)}
\item \Adef{W}{Supported by the Ministry of Science and Technology (Taiwan)}
\item \Adef{X}{Supported by the Israel Academy of Sciences and Humanities (Israel)}
\item \Adef{Y}{Supported by the Tomsk Polytechnic University Competitiveness Enhancement Program (Russia)}
\item \Adef{Z}{Supported by the National Science Foundation, Grant no. PHY-1506416 (USA)}
\item \Adef{a}{Supported by NCN, Grant 2017/26/M/ST2/00498 (Poland)}
\end{Authlist}

\clearpage
}
\setcounter{page}{1}

\section{Introduction}
\label{intro}

In this paper, exclusive  $\omega$ meson muoproduction is studied in the  process 
\begin{equation}
\mu + p \rightarrow \mu' + p' + \omega,
\label{eq:introduction:01}
\end{equation}
which in the one-photon-exchange approximation
 is described by the interaction of a virtual photon $\gamma^{*}$
%emitted from the incoming muon with one of the partons from
with the target proton $p$:
\begin{equation}
\gamma^{*} + p \rightarrow p' + \omega. 
\label{eq:introduction:02}
\end{equation}
This process, which at high virtuality $Q^2$ of the photon is known as Hard Exclusive Meson Production (HEMP), serves at low values of the squared four-momentum transfer $t$ as an important tool to access 
Generalised Parton Distributions (GPDs) \cite{gpd1, gpd2, gpd3, gpd4,
Radyushkin:1996ru} that contain a wealth of new information on the parton structure of the nucleon.

The amplitude for Hard Exclusive Meson Production by
longitudinally polarised virtual photons was proven to factorise into a
hard-scattering part, which
%that 
is calculable in perturbative QCD (pQCD),  
and a soft part~\cite{gpd4, Collins:1996fb}. The soft part contains 
GPDs that describe the structure of the probed nucleon and a
distribution amplitude %, which 
that accounts for the structure
of the produced meson. The factorisation is referred to as collinear
because parton transverse momenta are neglected. No similar proof of
collinear factorisation exists for transversely polarised virtual
photons. However, phenomenological
pQCD-inspired models have been proposed~\cite{Martin-1997,
Goloskokov:2005,Goloskokov:2008,Goloskokov:2009}
that go beyond the collinear factorisation by postulating the so
called $k_{\perp}$ factorisation, where
$k_{\perp}$ denotes parton transverse momentum.
In the Goloskokov-Kroll model~\cite{Goloskokov:2005,Goloskokov:2008,
Goloskokov:2009,GK:epjC-2014,GK:epjA-2014},
hereafter referred to as GK model, cross sections, Spin Density Matrix Elements (SDMEs) 
as well as 
target %-
and beam-spin asymmetries for HEMP by both   longitudinally and transversely polarised
virtual photons can be described simultaneously.

% At leading twist, vector-meson production by longitudinal virtual photons is described by the 
% chiral-even GPDs 
% $H^{f}$ and $E^{f}$, 
% where $f$ denotes a quark of a given flavour or a gluon. When higher-twist
% effects are included in the three-dimensional {\blue {[PK:] light-cone}} meson wave function,
% in addition to chiral-even GPDs $H^{f}$, $E^{f}$, $\widetilde{H}^{f}$ and $\widetilde{E}^{f}$, the chiral-odd GPDs $H_T^{f}$ and $\bar{E}_T^{f}$ appear, which describe process
% amplitudes with helicity flip of the `active' quark.

At leading twist, longitudinally polarised vector-meson production by longitudinally polarised virtual
photons is described by the chiral-even GPDs $H^{f}$ and $E^{f}$, where $f$ denotes a quark of a given flavour
or a gluon.
When higher-twist effects are included in the three-dimensional light-cone wave function, 
the production of 
longitudinally polarised
vector mesons by transversely polarised virtual photons  is described  by  the chiral-odd
GPDs $H_T^{f}$ and $\bar{E}_T^{f}$, which allow a helicity flip of the ``active'' quark.
Unnatural-parity exchange (UPE) contributes also to transitions 
from transversely polarised virtual photons to transversely polarised vector mesons or
from longitudinally polarised virtual photons to transversely polarised vector mesons.
These contributions are described by the GPDs $\widetilde{H}^{f}$ and $\widetilde{E}^{f}$.
Besides these UPE contributions, there is a sizeable pion-pole contribution %, which 
that
is treated as a one-boson exchange in the GK model.

 The SDMEs describe the spin structure of the reaction shown in Eq.~(\ref{eq:introduction:01}).
They are related to helicity amplitudes that describe transitions between specified spin states of virtual
 photon, target proton, produced vector meson and recoil proton. For an unpolarised nucleon target, after summing over initial and final spin states of the proton, SDMEs only depend on the helicities of virtual photon and produced meson.
 The measured SDME values can be used to establish a hierarchy of helicity amplitudes, 
 to test the hypothesis of $s$-channel helicity conservation (SCHC), 
to evaluate the contribution of unnatural-parity-exchange transitions and to assess the role of  
chiral-odd, 
i.e.\ parton helicity-flip GPDs 
in exclusive $\omega$ production. They  also allow 
to determine the phase difference between helicity amplitudes as well as the longitudinal-to-trans\-ver\-se cross-section ratio. 
The measurements of SDMEs can provide further constraints on GPD
parameterisations
beyond those from measurements of cross sections and spin asymmetries
for HEMP.

The HERMES measurements of SDMEs for exclusive electroproduction of $\omega$ mesons 
\cite{HERMES:2014} in the kinematic region 1.0~(GeV/$c$)$^2$ $< Q^{2} <$ 10~(GeV/$c$)$^2$, 3.0~GeV/$c^2$ $< W < 6.3$~GeV/$c^2$
and $|t| < 0.2$~(GeV/$c$)$^2$,
where $t$ is the squared
four-momentum transfer to the target,
indicate a sizeable contribution of UPE transitions that
 can be described by GPDs $\widetilde{H}^{f}$ and $\widetilde{E}^{f}$ 
 related to quark helicity
distributions. 
Here, $Q^2$ denotes the virtuality of the exchanged virtual photon and $W$ is the mass of the final hadronic system.
In the framework of the GK model
it turns out \cite{GK:epjA-2014} that the pion-pole exchange, which is treated as one-boson exchange, is an important contribution required to reproduce the HERMES results. The effect of such a $t$-channel $\pi^0$ exchange decreases with $W$
while it is predicted still to be measurable at COMPASS. The HERMES results on
SDMEs for exclusive $\omega$ production, as well as those for exclusive
$\rho^{0}$ production \cite{HERMES:2014, DC-24}, indicate a violation of the SCHC hypothesis, which in
the framework of the GK model is attributed to a contribution of chiral-odd GPDs. 

Also, an early paper on exclusive $\omega$  electroproduction~\cite{joos} contains results on SDMEs obtained
at DESY for 0.3~(GeV/$c$)$^2$ $< Q^{2} < 1.4$~(GeV/$c$)$^2$ and 0.3~GeV/$c^2$ $< W < 2.8$~GeV/$c^2$.
The SDMEs in exclusive $\omega $ electroproduction  were also
studied~\cite{clas} at CLAS in the range 1.6~(GeV/$c$)$^2$ $< Q^2 <$
5.2~(GeV/$c$)$^2$ and 1.9~GeV/$c^2$ $< W <$
2.8~GeV/$c^2$. It was  found that the exchange of the pion Regge
trajectory dominates exclusive $\omega$ production, even for $Q^2$ values
as large as 5~(GeV/$c$)$^2$.
% {\color{red} At the highest energies exclusive $\omega$ electroproduction has been studied by ZEUS at HERA~\cite{zeus} in the kinematic region 
% 3~(GeV/$c$)$^2$ $< Q^{2} <$ 20~(GeV/$c$)$^2$, 40~GeV/$c^2$ $< W < 120$~GeV/$c^2$
% and $|t|$ < 0.6~(GeV/$c$)$^2$. 
% Albeit extraction of SDMEs was not feasible because of limited statistic, the cross sections were measured as a function of $W$ and $Q^{2}$.} 
 
The present COMPASS results on SDMEs for exclusive $\omega$ muoproduction, 
which supplement the published COMPASS results on azimuthal asymmetries for 
transversely polarised protons \cite{COMPASS-asy}, have the potential to
further constrain GPDs. In the framework of the GK model it may become possible 
to assess the role of chiral-odd GPDs in the mechanism  
of SCHC violation  
and to shed light onto 
the mechanism of UPE transitions. 

\section{ Theoretical formalism}
\label{formalism}

Adopting the notation from Ref.~\cite{HERMES:2014}, the theoretical formalism of SDMEs and helicity amplitudes introduced by K. Schilling and 
G. Wolf~\cite{Schill} is used throughout this paper.

\subsection{Definition of Spin Density Matrix Elements}
 
The  helicity amplitudes {$F_{\lambda _{V} \lambda '_{N}\lambda _{\gamma} \lambda
_{N}}$} describe the transition of a
virtual photon with helicity $\lambda _{\gamma}$ to a vector meson with  
helicity $ \lambda _{V}$,
where $\lambda_{N}$ ($\lambda '_{N}$) is the
nucleon helicity  in the initial (final) state.
The helicity amplitudes depend on $W$, $Q^{2}$, and $t'\equiv|t| - t_{0}\approx p^{2}_{\rm T}$, where $t_{0}$ represents the smallest kinematically allowed value of $|t|$ for given
$Q^{2}$ and meson mass, and $p^{2}_{\rm T}$ is the square of the vector-meson transverse
momentum with respect to the direction of the virtual photon. In the centre-of-mass (CM) system of virtual photon and nucleon, the vector-meson spin density matrix $\rho_{\lambda_{V}\lambda_{V}^{'}}$ is related to the helicity amplitude {$F_{\lambda _{V} \lambda '_{N}\lambda _{\gamma} \lambda_{N}}$} as
%given by equation~\ref{eq:rhotohel} 
\cite{Schill}
\begin{eqnarray}
\rho_{\lambda_{V} \lambda '_{V}}=   
\frac{1}{2 \mathcal{N} }
  \sum_{\lambda_{\gamma}
\lambda '_{\gamma}\lambda_N \lambda '_N}
   F_{\lambda_{V}\lambda '_N\lambda_{\gamma}\lambda _N}~
 \varrho^{U+L}_{\lambda_{\gamma} \lambda '_{\gamma }}~   
  F_{\lambda '_{V} \lambda '_N\lambda '_{\gamma}\lambda
 _N}^{*},\,\,
 \label{eq:rhotohel}
\end{eqnarray}
where $\mathcal{N}$ is a normalisation factor~\cite{Schill,DC-24}. 
%\sout{the $\varrho^{U+L}_{\lambda_{\gamma} \lambda'_{\gamma}} $ is the virtual photon spin density matrix~\cite{DC-24}.} 
The virtual-photon spin density matrix 
$\varrho^{U+L}_{\lambda_{\gamma} \lambda'_{\gamma}}$  ~\cite{DC-24} describes the subprocess
$ \mu \rightarrow \mu' + \gamma^{*}$, which is 
calculable in quantum electrodynamics. It can be decomposed as 
\begin{equation}
\varrho^{U+L}_{\lambda_{\gamma} \lambda '_{\gamma }}
 = \varrho^{U}_{\lambda_{\gamma} \lambda '_{\gamma }} +
P_{b}~\varrho^{L}_{\lambda_{\gamma} \lambda '_{\gamma }},
\label{phspden}
\end{equation}
where the matrix with superscript $L$ ($U$) contains elements that
are coupled (not coupled) to the beam polarisation %and 
$P_{b}$. % is the value of the beam polarisation.
In the following the corresponding vector-meson SDMEs, which are related to these elements, will be referred to as ``polarised'' (``unpolarised'').

 %\sout{After decomposition of %$\varrho^{U+L}_{\lambda_{\gamma} \lambda
 %'_{\gamma}} $ into the standard  set of $3\times 3$ %Hermitian
 %matrices $ \Sigma^{\alpha}$, 
 %the vector-meson spin density matrix
 %is expressed in 
 %terms of a set of nine matrices %$\rho^{\alpha}_{\lambda_V\lambda'_V}$, which are %related to various 
 %photon-polarisation states: transversely polarised %photon ($\alpha$=0, ... ,3), 
 %longitudinally  polarised photon ($\alpha$=4), and  %terms describing  
 %their interference  ($\alpha$=5, ..., %8)~\cite{Schill}.}
The vector-meson spin density matrix   %$\varrho^{U+L}_{\lambda_{\gamma} \lambda
 %'_{\gamma}}$ 
 can be decomposed into a set of nine matrices
 $\rho^{\alpha}_{\lambda_V\lambda'_V}$  corresponding to different virtual-photon polarisation states: transverse polarisation ($\alpha$=0, ..., 3), 
 longitudinal polarisation ($\alpha$=4), and  
 their interference  ($\alpha$=5, ..., 8)~\cite{Schill}.
%In case it is 
If it is experimentally %] 
not possible to separate the contributions from longitudinally and transversely polarised photons, 
%it is usual to define the SDMEs as
SDMEs are usually defined as follows:
\begin{equation}
r^{04}_{\lambda_{V}\lambda '_{V}} = (\rho^{0}_{\lambda_{V}\lambda '_{V}}
+ \epsilon R \rho^{4}_{\lambda_{V}\lambda '_{V}})( 1 + \epsilon R )^{-1},   
\nonumber
\end{equation}
\begin{equation}
r^{\alpha}_{\lambda_{V}\lambda'_{V}} =
\begin{cases}
{  \rho^{\alpha}_{\lambda_{V}\lambda'_{V}}}{(  1 + \epsilon R )^{-1}},
\; \alpha = 1,2,3,\\
{ \sqrt{R} \rho^{\alpha}_{\lambda_{V}\lambda '_{V}}}
{(1 + \epsilon R )^{-1}},\; \alpha = 5,6,7,8.
\end{cases}
  \hspace*{0.25cm}
\label{rmatr}
\end{equation}

Here, $R= d\sigma_{L}/ d\sigma_{T}$ is the differential longitudinal-to-transverse cross-section ratio of virtual photons and $\epsilon$ is the
virtual-photon polarisation parameter, see Eq.~(\ref{expreps}).
The relations between the 23 SDMEs defined in Eq.~(\ref{rmatr})
% \magenta{[WDN: we better use all indices here: $r^{\alpha}_{\lambda_{V}\lambda'_{V}}$}]
% $r$ 
and the helicity amplitudes are given in Appendix A of Ref.~\cite{DC-24}.

\subsection{Properties of Helicity Amplitudes}
\label{hel_ampli}

 \indent As detailed in Refs.~\cite{Schill,DC-24}, each helicity amplitude can be linearly decomposed into a natural-parity-exchange (NPE) amplitude $T$ and an unnatural-parity-exchange (UPE) amplitude $U$,
\begin{equation}
F_{\lambda_{V} \lambda '_{N} \lambda_{\gamma}  \lambda_{N} } =
T_{\lambda_{V} \lambda '_{N} \lambda_{\gamma}  \lambda_{N} }+
U_{\lambda_{V}\lambda '_{N} \lambda_{\gamma}  \lambda_{N}},
\label{nu}
\end{equation} 

with the following relations~\cite{Schill}:
\begin{eqnarray}
T_{\lambda_V \lambda'_N \lambda_{\gamma} \lambda_N}=\frac{1}{2}[
F_{\lambda_V \lambda'_N \lambda_{\gamma} \lambda_N}
%\nonumber ~~~~~~~~~~~~~~~~~~~~~~~~~~~\\
+(-1)^{\lambda_V-\lambda_{\gamma}}F_{-\lambda_V \lambda'_N -\lambda_{\gamma}\lambda_N}],
\label{fnat}\\
U_{\lambda_V \lambda'_N \lambda_{\gamma} \lambda_N}=\frac{1}{2}[
F_{\lambda_V \lambda'_N \lambda_{\gamma} \lambda_N} 
%\nonumber~~~~~~~~~~~~~~~~~~~~~~~~~~~ \\
-(-1)^{\lambda_V-\lambda_{\gamma}}F_{-\lambda_V \lambda'_N -\lambda_{\gamma} \lambda_N}].
\label{funnnat}
\end{eqnarray}
% \red{\sout{The asymptotic behaviour of amplitudes $F$
% at small $t'$~\cite{Diehl},}}
% \red{\begin{eqnarray}
% F_{\lambda_V \lambda'_N \lambda_{\gamma} \lambda_N} \propto \Bigl (\frac{\sqrt{t'}}{M}\Bigr
% )^{|(\lambda_V-\lambda'_N)-(\lambda_{\gamma}-\lambda_N)|},
% \label{asytpr}
% \end{eqnarray}
% }

% \red{\sout{follows from  angular-momentum conservation. Here and in the following $M$ denotes 
% the proton mass. 
% Equations (\ref{fnat}-\ref{asytpr}) show that double-helicity-flip amplitudes with 
% $|\lambda _V-\lambda_{\gamma}|=2$ are suppressed at least by a factor of $\sqrt{t'}/M$, and that their contributions to SDMEs are suppressed by $t'/M^2$.}} 

% \blue{There is no interference between NPE and UPE amplitudes and no linear contribution from nucleon-helicity-flip amplitudes to SDMEs in case of an unpolarised target.}\blue{\bf [$\leftarrow$ This sentence is a bit left alone and it is not clear to me what use it has here?]}
Using the notation
\begin{eqnarray}
\widetilde{\sum}T_{\lambda_V \lambda_{\gamma}} T^*_{\lambda'_V
\lambda'_{\gamma}}\equiv 
 \frac{1}{2} \sum_{\lambda_N \lambda'_N}
T_{\lambda_V \lambda'_N\lambda_{\gamma}\lambda_N}
T^*_{\lambda'_V \lambda'_N\lambda'_{\gamma}\lambda_N}.
\label{tilde-sum}
\end{eqnarray}

and the symmetry properties~\cite{Schill,DC-24} of the amplitudes
$T$, Eq.~(\ref{tilde-sum}) becomes
\begin{eqnarray}
\widetilde{\sum}T_{\lambda_V \lambda_{\gamma}} T^*_{\lambda'_V
\lambda'_{\gamma}}= 
T_{\lambda_V \frac{1}{2}\lambda_{\gamma}\frac{1}{2}}
T^*_{\lambda'_V \frac{1}{2}\lambda'_{\gamma}\frac{1}{2}}+
T_{\lambda_V -\frac{1}{2}\lambda_{\gamma}\frac{1}{2}}
T^*_{\lambda'_V -\frac{1}{2}\lambda'_{\gamma}\frac{1}{2}}.~~~~~
\label{sum-two}
\end{eqnarray}
Here, both products on the right-hand side represent the contribution of 
NPE amplitudes, the first without and the second with nucleon-helicity flip. The relations for the UPE amplitudes can be written in an analogous way.
In the abbreviated notation used in the text, the nucleon-helicity indices will be omitted
for amplitudes with $\lambda_N=\lambda'_N$, i.e.
\begin{eqnarray}
T_{\lambda_V \lambda_{\gamma}}
&\equiv T_{\lambda_V \frac{1}{2}\lambda_{\gamma}\frac{1}{2}}&=
\phantom{-}T_{\lambda_V -\frac{1}{2}\lambda_{\gamma}-\frac{1}{2}},\nonumber\\  
 U_{\lambda_V \lambda_{\gamma}}
&\equiv U_{\lambda_V \frac{1}{2}\lambda_{\gamma}\frac{1}{2}}
&=-U_{\lambda_V -\frac{1}{2}\lambda_{\gamma}-\frac{1}{2}}.
\label{abrr}
\end{eqnarray}

The hypothesis of $s$-channel helicity conservation implies that there exist only
%, there exist only
%The dominance of
diagonal $\gamma^{*} \to V$ transitions ($\lambda_{V}=\lambda_{\gamma}$).
%is referred to as $s$-channel helicity conservation.

%\newpage
\section{Experimental access to SDMEs}
\label{access}

Spin density matrix elements are extracted from experimental data on exclusive muoproduction of $\omega$ mesons.
The SDMEs are fitted as parameters of the three-di\-men\-sio\-nal angular distribution 
$\mathcal{W}^{U+L}(\Phi,\phi,$ 
$\cos\Theta)$
to the corresponding experimental distribution. 
Here, $\Phi$ is the azimuthal angle of the produced $\omega$ meson, while the polar angle $\Theta$ and the azimuthal angle $\phi$ describe the three-pion decay of the $\omega$ meson, see Eqs.~(\ref{phicap1-def} - \ref{phismall2-def}).
%As shown in Eq.~(\ref{eqang1}), t
 The angular distribution $\mathcal{W}^{U+L}$ is decomposed
into contributions that are not coupled ($\mathcal{W}^{U}$) or coupled ($\mathcal{W}^{L}$) to the beam polarisation:
\begin{eqnarray}
\mathcal{W}^{U+L}(\Phi,\phi,\cos{\Theta}) =  \mathcal{W}^{U}(\Phi,\phi,\cos{\Theta}) + 
P_{b}\mathcal{W}^{L}(\Phi,\phi,\cos{\Theta}).%, 
\label{eqang1} %\hspace*{6.8cm}
\end{eqnarray}
%from an unpolarised beam, $\mathcal{W}^{U}$, and a longitudinally polarised beam, $\mathcal{W}^{L}$.
%From the data that 
Using the data, which were collected  with a longitudinally
polarised beam, 15 ``unpolarised'' SDMEs %(see Eq.~(\ref{eqang2}))
are extracted from $\mathcal{W}^U$:
%and 8 ``polarised'' SDMEs 
%(see Eq.~(\ref{eqang3})) 
%are extracted:
%%%\begin{figure*}[hbt!]
%\begin{eqnarray}
%\mathcal{W}^{U+L}(\Phi,\phi,\cos{\Theta}) =  \mathcal{W}^{U}(\Phi,\phi,\cos{\Theta}) + 
%P_{b}\mathcal{W}^{L}(\Phi,\phi,\cos{\Theta}), 
%\label{eqang1} \hspace*{6.8cm}
%\end{eqnarray}
\begin{eqnarray}
\mathcal{W}^{U}(\Phi,\phi,\cos{\Theta})  &=& \frac{3} {8 \pi^{2}} \Bigg[
         \frac{1}{2} (1 - r^{04}_{00}) + \frac{1}{2} (3 r^{04}_{00}-1) \cos^2{\Theta}
\nonumber \\         
&-& \sqrt{2} \mathrm{Re} \{ r^{04}_{10} \} \sin 2\Theta 
\cos \phi 
- r^{04}_{1-1}  \sin ^{2} \Theta\cos 2 \phi %\hspace*{0.0cm}
\nonumber \\
&-& \epsilon \cos 2 \Phi \Big( r^{1}_{11} \sin ^{2} \Theta  + r^{1}_{00} \cos^{2}{\Theta}
  - \sqrt{ 2}  \mathrm{Re} \{r^{1}_{10}\} \sin 2  \Theta  \cos  \phi
    - r^{1}_{1-1} \sin ^{2} \Theta \cos 2 \phi   \Big)   \nonumber  \\
&-& \epsilon \sin 2 \Phi \Big( \sqrt{2} \mathrm{Im} \{r^{2}_{10}\} \sin 2 \Theta \sin \phi +
       \mathrm{Im} \{ r^{2}_{1-1} \} \sin ^{2} \Theta \sin 2 \phi  \Big)  \nonumber \\
&+& \sqrt{ 2 \epsilon (1+ \epsilon)}  \cos \Phi
\Big(  r^{5}_{11} \sin ^2 {\Theta} +
 r^{5}_{00} \cos ^{2} \Theta - \sqrt{2} \mathrm{Re} \{r^{5}_{10}\} \sin 2 \Theta \cos \phi \nonumber \\
&-& r^{5}_{1-1} \sin ^{2} \Theta \cos 2 \phi  \Big)  \nonumber \\
&+& \sqrt{ 2 \epsilon (1+ \epsilon)}  \sin \Phi
\Big( \sqrt{ 2} \mathrm{Im} \{ r^{6}_{10} \} \sin 2 \Theta \sin \phi
+ \mathrm{Im} \{r^{6}_{1-1} \} \sin ^{2} \Theta \sin 2 \phi \Big) \Bigg],
\label{eqang2}%\\ 
%\hspace*{-3.0cm}
\end{eqnarray}
and 8 ``polarised'' SDMEs from $\mathcal{W}^L$:
\begin{eqnarray}
\mathcal{W}^{L}(\Phi,\phi,\cos \Theta)  &=& \frac{3}{8 \pi^{2}}  \Bigg[
  \sqrt{ 1 - \epsilon ^{2} }  \Big(  \sqrt{ 2}  \mathrm{Im} \{ r^{3}_{10} \}
\sin 2 \Theta \sin \phi +
   \mathrm{Im} \{ r^{3}_{1-1}\} \sin ^{2} \Theta \sin 2 \phi  \Big)  \nonumber  \\
&+& \sqrt{ 2 \epsilon (1 - \epsilon)} \cos \Phi
\Big( \sqrt{2} \mathrm{Im} \{r^{7}_{10}\} \sin 2 \Theta \sin \phi
+  \mathrm{Im} \{ r^{7}_{1-1} \}  \sin ^{2} \Theta \sin 2 \phi   \Big)  \nonumber \\
&+& \sqrt{ 2 \epsilon (1 - \epsilon)} \sin \Phi
\Big( r^{8}_{11} \sin ^{2} \Theta + r^{8}_{00} \cos ^{2}
\Theta -  \sqrt{2} \mathrm{Re}\{ r^{8}_{10}\} \sin 2 \Theta \cos \phi   \nonumber \\ 
&-& r^{8}_{1-1} \sin ^{2} \Theta \cos 2\phi \Big)  \Bigg].\hspace*{-20.0cm} 
\label{eqang3}
\end{eqnarray}
%%%\end{figure*}

%T
Here, the virtual-photon polarisation parameter $\epsilon$
represents the  ratio of fluxes of
longitudinal and transverse virtual photons,
\begin{eqnarray}\label{expreps}
\epsilon &=& \frac{1-y - y^2\frac{Q^2}{4\nu^2}}{1-y+ \frac{1}{4}y^2
(\frac{Q^2}{\nu^2} + 2)}, 
\end{eqnarray}
where $y = p\cdot q / p\cdot k   \stackrel{lab}{=} \nu / E$.
%Here 
The symbols $p$, $q$ and $k$ denote the four-momenta of target proton, virtual
photon and incident lepton respectively. The energy of virtual photon and incident lepton in the target rest frame is denoted by $\nu$ and $E$, respectively.
%\blue{\bf $\leftarrow$ This entire text around Eq. 12 including the equation itself could be skipped. Epsilon is anyway introduced earlier (line 215) without defining it at first; is it really necessary in a paper that just briefly repeats the formalism to define it using a full paragraph? If the paragraph is removed, also the reference to Eq. 12 has to be removed in line 215. PS: the HERMES paper that a few lines were copied from does NOT define epsilon.}

\begin{figure}[hbt!]
\begin{center}
\includegraphics[width=8.0cm]{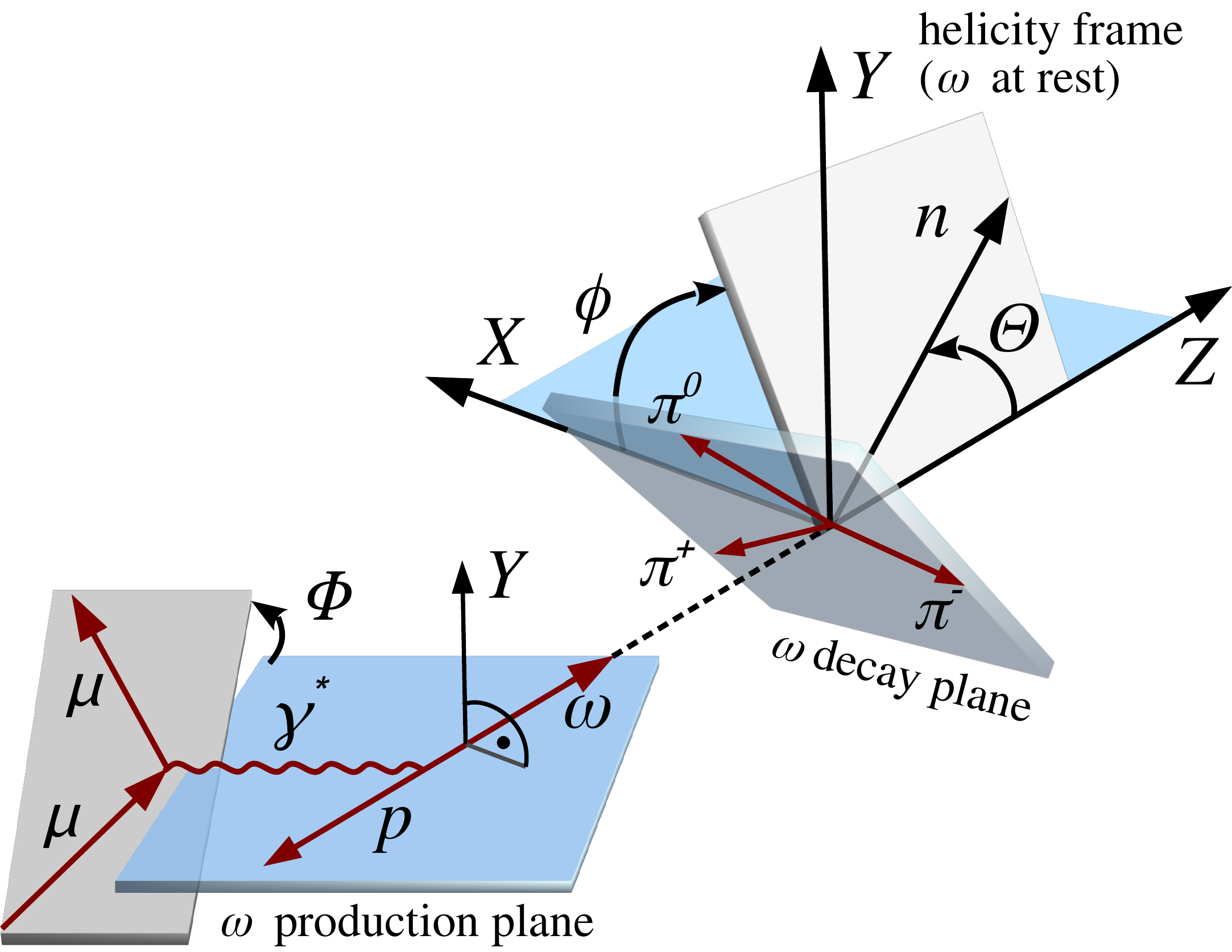}
%\vspace{0.5cm}   
\caption{
Definition of angles in the process
$\mu N \to \mu N  \omega$ with  $\omega 
 \to \pi^+ \pi^- \pi^0$. 
Here, $\Phi$ is the angle between the $\omega$ production plane
and the lepton scattering plane  in the
centre-of-mass system
of the virtual photon and the target nucleon.
The variables $\Theta$ and $\phi$ 
are respectively the polar and azimuthal
angles of the unit vector normal to the decay plane in the
$\omega$ meson rest frame.
}
\label{defang}
\end{center}
\end{figure}

Angles and reference frames are defined in Fig.~\ref{defang}.

The directions of axes of the ``hadronic CM system'' and 
the $\omega$-meson rest frame coincide with the directions of axes of the helicity
 frame~\cite{Schill,DC-24,joos}. Following Ref.~\cite{Schill}, the right-handed
hadronic CM system
of virtual photon and 
target nucleon, with coordinates $XYZ$,
is defined such that
the $Z$-axis  is aligned along the virtual-photon three-momentum $\vec{q}$
and the $Y$-axis is parallel to $\vec{q} \times \vec{v}$,  where $\vec{v}$
is the three-momentum of the $\omega$ meson.

For the convenience of the reader, we give in the following the explicit definitions of angles \cite{HERMES:2014}. 
The angle  $\Phi$   
between $\omega $ production plane 
and  lepton scattering plane  in the hadronic CM system 
is given  by 
\begin{eqnarray}
\cos \Phi = \frac{ (\vec{q} \times \vec{v}) \cdot (\vec{k} \times \vec{k}')}
{ | \vec{q} \times \vec{v} | \cdot |\vec{k} \times \vec{k}'| } 
\label{phicap1-def}
\end{eqnarray}
and
\begin{eqnarray}
\sin \Phi = 
 \frac{ [ (\vec{q} \times \vec{v} )\times (\vec{k} \times \vec{k}' )] \cdot \vec{q} }
 { |\vec{q} \times \vec{v}| \cdot |\vec{k} \times \vec{k}'| \cdot |\vec{q}|
 }.
\label{phicap2-def}
\end{eqnarray}    
Here  $\vec{k}$, $\vec{k'}$, $\vec{q}=\vec{k}-\vec{k'}$ and  $\vec{v}$  are the three-momenta of the incoming  and outgoing
lepton, the virtual photon
and the $\omega $ meson respectively.\\
\indent The unit vector  normal  to the decay plane in the $\omega$ rest frame is defined by 
\begin{eqnarray}
\vec n= \frac{\vec{p}_{\pi^+} \times \vec{p}_{\pi^-}}{|\vec{p}_{\pi^+}
\times\vec{p}_{\pi^-}| },
\label{normal}
\end{eqnarray}
where $\vec{p}_{\pi^+}$ and  $\vec{p}_{\pi^-}$ are the three-momenta of the
positive and  negative  decay pion in the $\omega$ rest frame, respectively.

The polar angle  $\Theta$ of the unit vector  $\vec{n}$ in the 
$\omega$ meson rest frame, with the $z$-axis aligned opposite to the outgoing nucleon
momentum $\vec{p}'$  and the $y$-axis directed along
$\vec{p}' \times \vec{q}$, is defined by 
\begin{eqnarray} 
\cos \Theta  = -\frac{ \vec{p}' \cdot \vec n }{| \vec{p}' |}. 
\label{theta-def}
\end{eqnarray} 
The azimuthal angle $\phi$ of the unit vector $\vec n$ is given
as follows:
\begin{eqnarray} 
\cos \phi = 
\frac{ (\vec{q} \times \vec{p}' )\cdot (\vec{p}' \times \vec n) }
{ | \vec{q} \times \vec{p}'| \cdot |\vec{p}' \times \vec{n} | },
\label{phismall1-def}
\end{eqnarray}
\begin{eqnarray}
 \sin \phi = - 
 \frac{[ (\vec{q} \times \vec{p}' )\times \vec{p}' ] \cdot ( \vec n  \times \vec{p}' ) } 
{ | (\vec{q} \times \vec{p}' )\times \vec{p}' | \cdot |\vec n \times \vec{p}' |
 } .
\label{phismall2-def}
\end{eqnarray}

\section{Experimental setup and data selection}
\label{sec:exp} % CKR added

The main component of the COMPASS setup is the two-stage magnetic
 spectrometer. Each spectrometer stage comprises a dipole magnet
 complemented by a variety of tracking detectors, a muon filter for muon
 identification and an electromagnetic as well as a hadron
 calorimeter. A detailed description of the setup can be found in Refs.
 \cite{comp1,comp2,comp3}.

The data used for this analysis were collected within four weeks in 2012. In this period 
the COMPASS spectrometer was complemented by a 2.5 m long
liquid-hydrogen target surrounded by a time-of-flight (TOF) system for the detection of recoiling protons and by a
third electromagnetic calorimeter placed directly downstream of 
the target.
The recoil detector restricts the kinematic coverage towards the region of small squared transverse momentum of the $\omega$-meson with respect to the virtual-photon direction. Hence it was used only for an additional check of the background correction procedure as explained in Sec.~\ref{sec:sys} and not for the determination of SDMEs.

Data  with $\mu^{+}$ and $\mu^{-}$ beams were
 taken separately. The natural polarisation of the muon beam provided by the
 CERN SPS originates from the parity-violating decay-in-flight of the parent
 meson, which implies opposite polarisations for $\mu^{+}$ and $\mu^{-}$
 beams. Within regular time intervals during 
data taking, charge and
 polarisation of the muon beam were swapped simultaneously. In order to equalise the spectrometer acceptance for the two beam charges,
 also the polarities of the two spectrometer magnets were changed accordingly. For  both beams, 
the absolute value of the average beam polarisation is about 0.8 with an
 uncertainty of about 0.04.

An event to be accepted
for  analysis is required to have a topology as that of the
observed process
\begin{center}
\begin{tikzpicture}
\node [right] at (0., 1.2) {$\mu p \rightarrow \mu' p' \omega$};
\node [right] at (3.0, 0.6) {$\pi^+ \pi^- \pi^0$};
\node [right] at (5.1, 0.) {$\gamma \gamma$~~};
\draw [thick, ->] (2.1, 0.9) -- (2.1, 0.6) -- (2.7, 0.6);
\draw [thick, ->] (4.2, 0.3) -- (4.2, 0.) -- (4.8, 0.);
\node [right] at (6.7, 0.0) {$\mathrm{BR} \approx 99\%.$};
\node [right] at (6.7, 0.6) {$\mathrm{BR} \approx 89\%$};
\label{eq:selection:reaction}
\end{tikzpicture}
\end{center}
The selected events should have one reconstructed vertex inside the
liquid-hydrogen target  associated with the incoming and the outgoing muon, and
two hadron tracks of oppo\-si\-te charge. The outgoing muon has to have the
same charge as the incoming muon and is required to traverse more than 15 radiation lengths to be identified as a muon. The
charged hadron tracks are selected by requiring the traversed path to be
shorter than 10 radiation lengths.

\subsection{$\pi ^0$ reconstruction}
A neutral pion is reconstructed via its dominant decay into two photons that are
registered as neutral clusters  in the electromagnetic
calorimeters. As neutral cluster we denote a reconstructed
calorimeter cluster that is not associated to a charged track, thereby
including any cluster for the most upstream calorimeter %\st{that} \blue{, which} 
that had no
tracking system upstream of it.
The method of $\pi^0$ reconstruction
is  similar to  
the one used in  
the analysis of azimuthal asymmetries for
exclusive $\omega$ production on a trans\-versely polarised target ~\cite{psz}. 

\begin{figure}[hbt!]\centering
\includegraphics[width=8.0cm]{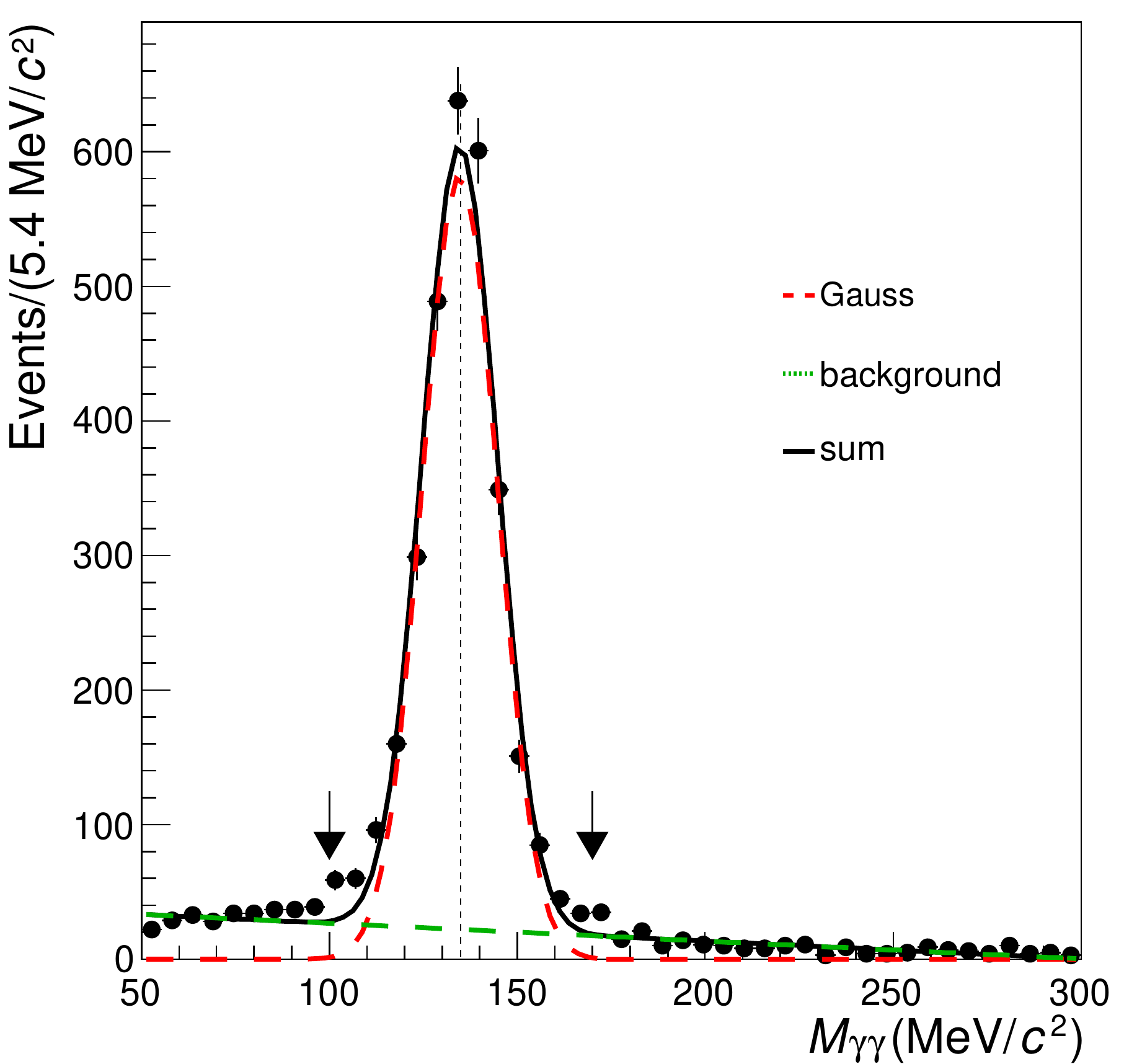}
\caption{\small{ 
Distribution of the two-photon invariant mass fitted by a Gaussian
 function and a linear background. The dashed vertical line denotes the
 PDG value of the $\pi^0$ mass.
The black arrows indicate the selection window.}}
\label{pi_1}
\end{figure}

In  Fig.~\ref{pi_1} the distribution of the reconstructed two-photon invariant
mass is shown. The $\pi ^{0}$
peak is prominent. 
The  distribution is fitted by a superposition of the signal, which is
described by a Gaussian function, and a linear background. 
After selection of an event with a $\pi^{0}$, the energies of the decay photons are scaled
 by the  factor $ M^{\mathrm{PDG}}_{\pi^{0}}/M_{\gamma\gamma}$ 
 in order to improve the experimental resolution of the reconstructed three-pion invariant mass. Here $
 M^{\mathrm{PDG}}_{\pi^{0}}\approx$ 0.135~$\mathrm{GeV/{\it c}}^2$ is the nominal 
 $\pi^0$ mass. This
 scaling does not  affect the angular distributions of 
 neutral pions.
 %, albeit it
 %improves the experimental resolution of the %reconstructed %3
 %three-pion invariant mass.  

\subsection{Kinematic selections}
\label{sec:kine-sel} % CKR added
\begin{figure}[hbt!]\centering
\includegraphics[width=7.5cm]{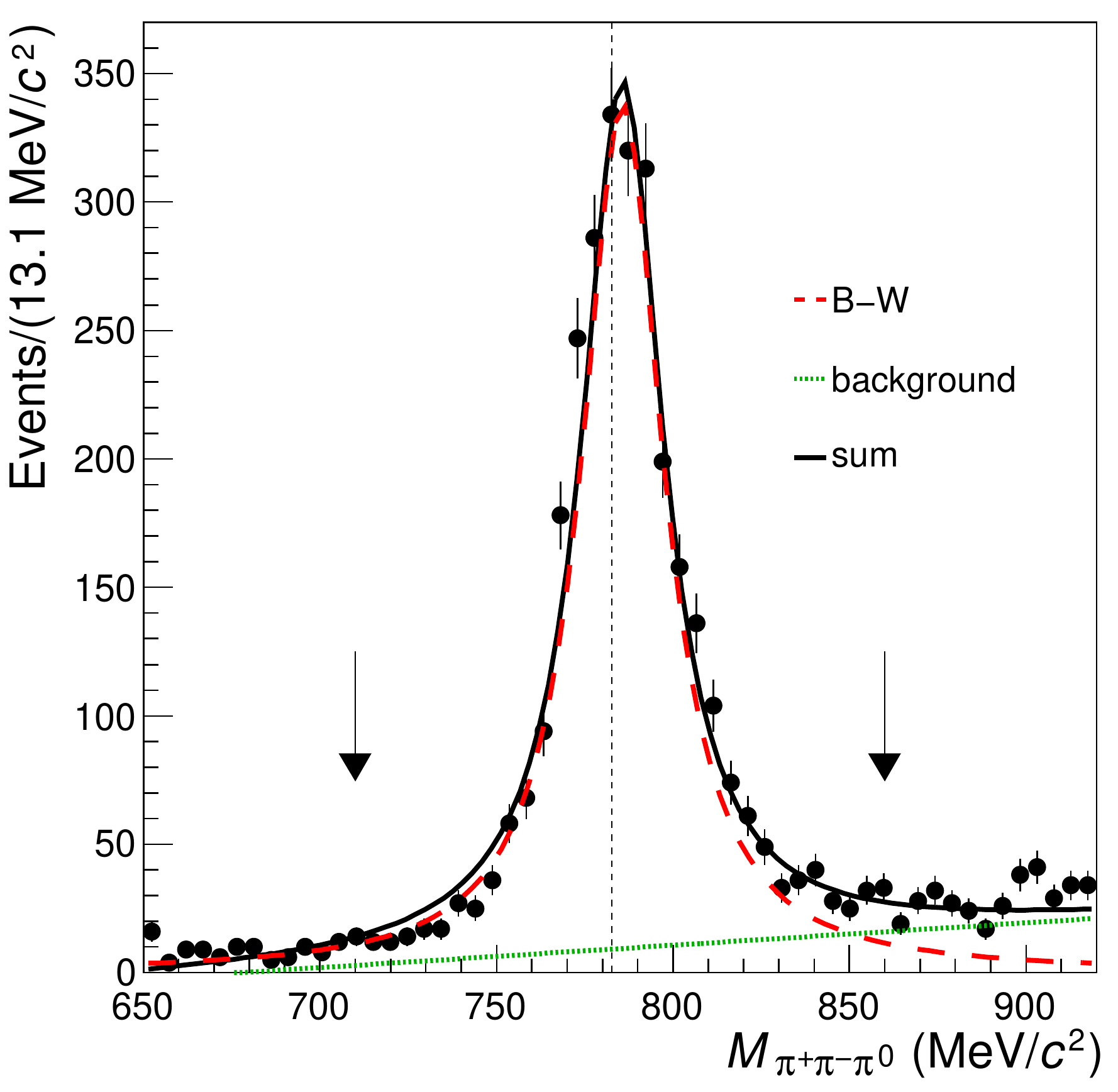}
\caption{\small{ 
Distribution of the $\pi ^+ \pi ^- \pi ^0$ invariant mass
 fitted by a Breit-Wigner function and linear background. The dashed
vertical line denotes the PDG value of the $\omega$ mass. The black arrows denote the applied limits.
 }}
\label{pi_om}
\end{figure}

The following kinematic selections
are applied to  
 %\st{select} \blue{extract}  \st{the} 
select exclusively produced
$\omega$ mesons:

\begin{itemize}
\item 1.0~(GeV/$\it{c})^2$ < $Q^{2} < 10.0$~(GeV/$\it{c})^2$, where the lower limit 
ensures applicability of pQCD and the upper one suppresses background due to hadrons produced in DIS, which hereafter is referred to as ``SIDIS background''.

\item 0.1 < y < 0.9, where the lower limit suppresses events with  poorly
reconstructed
kinematics and the upper one removes events with large radiative corrections.

\item $W$ > 5.0~GeV/$\it{c}^2$ to remove the kinematic region where the cross section changes rapidly due to the production of
resonances.

\item   0.01~(GeV/$\it{c})^2$  < $ p_{\rm{T}}^2 <  0.5$~(GeV/$\it{c})^2$, where $p_{\rm{T}}$ is the
transverse momentum of the $\omega$ meson  with
respect to the virtual-photon direction. 
The lower limit removes
events with a poorly determined
azimuthal angle of the produced meson and the upper one suppresses SIDIS background.

\item 0.1~GeV/$\it {c}^2 < M_{\gamma \gamma} < $0.17~GeV/$\it {c}^2$, where $M_{\gamma
\gamma}$ is the two-photon invariant mass, in order to select $\pi^0$ mesons.

\item 0.71~GeV/$\it{c}^2$ < $M_{\pi^{+}\pi^{-}\pi^{0}}$ <  0.86~GeV/$\it {c}^2$, where
$M_{\pi^{+}\pi^{-}\pi^{0}}$
is the three-pion invariant mass, in order to select $\omega$ mesons.
In  Fig.~\ref{pi_om}, the $\omega$ signal is clearly visible above a small background. The invariant
mass of the three-pion system is fitted by a Breit-Wigner function and a linear
background. As the non-resonant background is found to be small, i.e. about 7\% for the used range of three-pion invariant mass, its effect on the extraction of SDMEs is neglected. 

\item $ E_{\omega}$ > 14~GeV to reduce the SIDIS background
contribution.
\end{itemize}

The information from the recoil proton detector is not used for the extraction of SDMEs. Instead, in order to enhance the fraction of events with exclusively produced $\omega$ mesons, the missing energy
\begin{equation}
E_{\rm miss}  = \frac{ M^{2}_{\rm X} - M^{2}}{2 M}
\end{equation}
is constrained by $-$3.0~GeV $<  E_{\rm miss} <$ 3.0~GeV 
to take into account the experimental resolution. Here $M$ is 
the proton mass, $ M^{2}_{\rm X}=({p} + {q}- {p}_{\pi^+} - {p}_{\pi^-} -{p}_{\pi^0})^{2}$ 
is the missing mass squared, and ${p}_{\pi^+}$, $
{p}_{\pi^-}$ and  ${p}_{\pi^0}$ are the four-momenta of the three
 pions.
The $E_{\rm miss}$ distribution for the experimental data is shown in
Fig.~\ref{emiss} as open histogram. The exclusive peak is apparent.

After applying all  kinematic selection criteria, 3060  events are available for further analysis.

\subsection{SIDIS background}
\label{sec:SIDIS-bg}

\begin{figure}[hbt!]\centering
\includegraphics[width=8.cm]{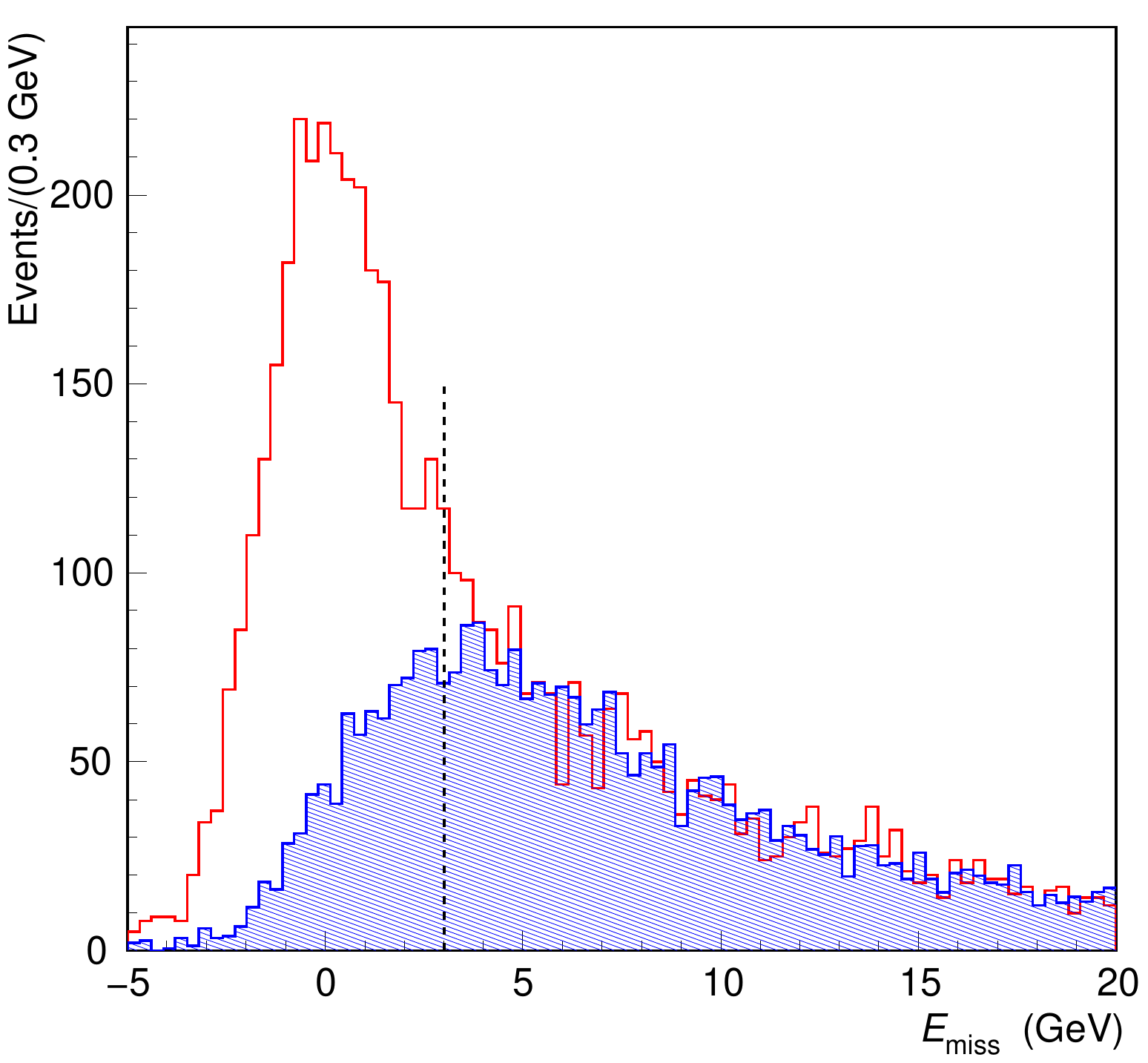}
\caption{ The missing-energy distribution from experimental data (open
histogram) compared to the distribution of SIDIS events from a LEPTO MC
simulation
(shaded histogram). The MC distribution is
normalised to the data in 
the region 7~GeV  $< E_{\rm miss} < 20$~GeV. The vertical dashed line denotes
the upper limit of the exclusive region. Each LEPTO MC event is
reweighted by a $E_{\rm miss}$-dependent weight that is calculated using both
experimental and simulated data 
with same-charge hadron pairs. See text for a detailed
explanation.  
 }
\label{emiss}
\end{figure}

The $ E_{\rm miss} $ distribution is used to determine the fraction of 
SIDIS background under the exclusive peak, following the procedure described
in Refs.~\cite{COMPASS-asy,psz}. For the simulation of background, the LEPTO 6.5.1 %Monte Carlo 
generator is used with the COMPASS tuning of parameters~\cite{Comptune}.
In order to achieve the best possible agreement between experimental and simulated 
$E_{\rm miss}$ distributions, the  
simulated data 
are reweighted on a bin-by-bin basis using the weight
\begin{equation}
 w(E_{\rm miss}) = 
 \frac{N^{sc}_{rd}(E_{\rm miss})}{N^{sc}_{MC}(E_{\rm miss})}.
\end{equation}
Here $N^{sc}_{rd}(E_{\rm miss})$ and $N^{sc}_{MC}(E_{\rm miss})$  are
numbers of events containing same-charge hadron pairs,
$h^{+} h^{+} \gamma\gamma$ and $h^{-} h^{-} \gamma\gamma$, in the %3
three-pion system for experimental and simulated data, respectively. In order to improve the statistical significance, the
constraint on the $\omega$ invariant mass is not used for the purpose of
estimating the weight $w$. 
The shaded histogram in Fig.~\ref{emiss} represents the simulated  
SIDIS background, which is generated by LEPTO 
and processed through the full simulation of the COMPASS setup~\cite{hepg3}, followed by the same event
reconstruction and selection procedure as for the real data, and then reweighted in the way
described above. The distribution is
normalised to the
experimental data  in the region
7~GeV $<E_{\rm miss}<$ 20~GeV. 
The fraction of background in the signal window $-$3.0~GeV < $E_{\rm miss}$ < 3.0~GeV 
%for the total kinematic range 
is found to be $f_{bg}$= 0.28 for the total kinematic range.
The fraction of SIDIS background increases with increasing 
$Q^2$ and 
$p^2_{\rm T}$, and it decreases with increasing $W$. For the results on kinematic
dependences of SDMEs, which are presented in the following, the background 
fraction $f_{bg}$ 
is evaluated separately for each kinematic bin, the values 
ranging 
between 0.20 
and 0.41.

\section{Extraction of SDMEs}
\subsection{Unbinned maximum likelihood method}
\label{sec:uml} % CKR added
The SDMEs are determined by an Unbinned Maximum Likelihood fit of the
function
${\mathcal{W}} ({\cal R};\Phi,\phi,\cos\Theta)$ to the experimental
three-dimensional
 angular distribution of
$\omega $ production and decay. The explicit expression for the dependence
 of ${\mathcal{W}}$
on SDMEs was given in Sec.~\ref{access} by Eqs.~(\ref{eqang1},
\ref{eqang2}, \ref{eqang3}). Here $\cal R$ denotes the set of 23 SDMEs $r
^{\alpha}_{\lambda_{V}{\lambda'_{V}}}$.
The negative log-likelihood function to be minimised reads
\begin{eqnarray}
-\ln L({\cal R})=
-\sum_{i=1}^{N}\ln\frac{\mathcal{W}^{U+L}({\cal
R};\Phi_{i},\phi_{i},\cos{\Theta_{i}})}{\widetilde{\mathcal
N}({\cal R})},
\label{loglik-def}
\end{eqnarray}
where $N$ is the number of selected events.The likelihood normalisation factor
\begin{equation}
 \widetilde {\mathcal N}({\cal R})=
\sum_{j=1}^{N_{MC}}\mathcal{W}^{U+L}({\cal
 R};\Phi_{j},\phi_{j},\cos{\Theta_{j}})
\label{loglik-def1}
\end{equation}
is calculated numerically using the sample of MC events generated with
the HEPGEN++ $\omega$ generator, in the following denoted by HEPGEN
\cite{hepg1,hepg2}. This
generator is used to model the kinematics of exclusive $\omega$ production. For
the purpose of the present analysis, the option with an isotropic
three-dimensional angular distribution of $\omega$ production and decay is
chosen. The generated events are passed through a complete description of the
COMPASS setup and the resulting  data are treated in the same
way as it was done for experimental data. The number of HEPGEN events is denoted $N_{MC}$
in Eq.~(\ref{loglik-def1}).

\subsection{Background-corrected SDMEs}
\label{sec:bg-corr-in}

In order to determine SDMEs that are corrected for SIDIS background, a two-step procedure is used. First, the parameterisation of the background angular distributions
is obtained by applying the above described
maximum likelihood method to selected SIDIS events simulated with the LEPTO generator. 
These events are required to pass the same selection criteria as experimental data. Performing an unbinned likelihood fit according to Eq.~(\ref{loglik-def}) using simulated events in the range $-$3.0~GeV$<E_{\rm miss} <$ 3.0~GeV yields 
the set $\cal B$ of 23 ``background SDMEs''.

\label{sec:bg-corr} % CKR added   
In the second step, the set $\cal B$ of background SDMEs %${\cal B}$ 
is used to extract the set $\cal R$ of background-corrected SDMEs %$\cal R$ 
by applying the unbinned maximum likelihood fit to the experimental data.
For this purpose the following negative log-likelihood function is
fitted:

\begin{eqnarray} 
%-\ln L({\cal R})=\nonumber
%~~~~~~~~~~~~~~~~~~~~~~~~~~~~~~~~~~~~~~~~~~~~~~~~~~~\\
-\ln L({\cal R})=
-\sum_{i=1}^{N}\ln\Bigl[ \frac{ (1-f_{bg})~\mathcal W^{U+L}({\cal
R};\Phi_{i},\phi_{i},\cos\Theta_{i})
}{\widetilde {\mathcal N}({\cal R},{\cal B})} %\nonumber\\
+\frac{f_{bg}~\mathcal W^{U+L}({\cal B}; \Phi_{i},\phi_{i}, \cos \Theta_{i})}
{\widetilde{\mathcal N}({\cal R},{\cal B})}\Bigr].
\label{logbac}
\end{eqnarray}
Here, $f_{bg}$ is the fraction of background events in the selected experimental data as determined in Sec.~\ref{sec:kine-sel} and $\widetilde{\mathcal N}$ is the
normalisation factor:
\begin{eqnarray}
\widetilde{\mathcal N}({\cal R},{\cal B})=
\sum_{j=1}^{N_{MC}}[(1-f_{bg})~\mathcal W^{U+L}({\cal R}; \Phi_{j},
\phi_{j},\cos\Theta_{j})%\nonumber\\
+f_{bg}~\mathcal W^{U+L}({\cal B}; \Phi_{j},
\phi_{j},\cos\Theta_{j})].
\label{logbacnor}
\end{eqnarray}

\subsection {Systematic uncertainties}
\label{sec:sys} % CKR added
The following sources of systematic uncertainties are considered:

\begin{itemize}
\item[i)]{\it Difference between results for $\mu ^+$ and $\mu ^-$ beams}\\
The $\mu^{+}$  beam intensity was about 2.7 times higher than that of the $\mu ^-$ beam. 
 A possible impact of this difference on the determination of SDMEs is  checked
by comparing the SDMEs extracted
separately for the
$\mu^{+}$ beam (negative polarisation) and the $\mu^{-}$
beam (positive polarisation).
For each SDME, half of the difference between the SDME values determined  with  opposite beam polarisations
 is taken as systematic uncertainty.

\item[ii)]{\it Influence of shifted $E_{\rm miss}$ peak position}\\
It was  observed in Ref.~\cite{comnote}  that certain SDME values depend on the position of the $E_{\rm miss}$ peak.
 The $E_{\rm miss}$ distribution shown
  in Fig.~\ref{emiss} is not precisely centred at zero, but slightly shifted towards negative values.
This results from an imbalance between the energy  
measured for 
  the incoming muon and the energies of the final-state particles measured in the forward spectrometer.
The effect of this  shift  on the extracted SDMEs is investigated by applying the
 small kinematic correction (+0.7~GeV/$c$) to the beam momentum that is
 needed to centre the $ E_{\rm miss} $ peak at zero. The difference between the
 values of final SDMEs and those obtained with corrected kinematics is taken as
 %\blue{the} 
 systematic uncertainty.
 
 A similar shift of the $E_{\rm miss}$ peak
 is obtained by rescaling the momenta of the final-state particles measured in the spectrometer. The differences between SDME values obtained without and with the rescaling are comparable to those obtained with corrected beam momentum. In order to avoid double counting, only the differences obtained with corrected beam momentum are taken as systematic uncertainties. 
 
\item[iii)]{\it Effect of background subtraction}\\
 As detailed in Sec.~\ref{sec:bg-corr}, 
 %and\ref{sec:bg-corr}}
  the background-corrected SDMEs are obtained with  background SDMEs  obtained from LEPTO events in the exclusive region 
  $-$3.0~GeV $< E_{\rm{miss}} <$ 3.0~GeV. 
As the angular distributions from LEPTO were never experimentally 
verified for the event selection used in the present analysis,
as a check
an alternative method  
is used, in which background SDMEs 
are estimated from experimental 
 data in the
region  7.0~GeV$< E_{\rm miss} <$ 20.0~GeV. 
The difference between SDMEs obtained by these two methods is taken as
 systematic uncertainty.
 
In addition, the
procedure for background correction  
is checked by using 
the data from the recoil-proton detector (RPD).
These data allow 
us to apply additional
selection  criteria on exclusive events~\cite{DVCSpaper, pi0paper}, which lead to  
a reduction of the non-exclusive
background by a factor of about 10. 
As a limited $p^2_{\rm T}$-range is covered by the RPD, %recoil-proton detector, 
the same limited kinematic region 
is used to compare the SDMEs %results 
obtained with and without %the 
RPD.
%recoil-proton detector.
The results %SDMEs extracted by the two methods
are consistent within statistical
uncertainties, hence no systematic uncertainty is assigned here.

\item[iv)]{\it Comparison of unbinned and  binned maximum
likelihood methods}\\
The two fitting methods are expected to yield consistent results for sufficiently large statistics. In this analysis however, when using the unbinned method the background treatment is different from that when using the binned method. In the former case, the angular dependence of the background is parameterised, while in the latter case the background is subtracted in each angular bin on a bin-by-bin basis. Comparing the results from the two methods hence probes a possible systematic uncertainty due to the background-correction procedure.
For each SDME, the systematic uncertainty is given by the difference
between the final value as obtained using the unbinned method and the
value obtained with the binned method.

\item[v)]{\it Sensitivity to the shapes of the kinematic distributions generated by HEPGEN}\\
As no experimental data exists
on the differential cross section for exclusive $\omega$ production at COMPASS energies, a model is used to simulate the process in HEPGEN.
In order to check the sensitivity of SDMEs to the shapes of kinematic distributions in the HEPGEN generator, the SDME extraction
was repeated by reweighting the MC events with weights depending on $Q^2$ and $\nu $. The weights 
are tuned such that the $Q^2$ and $\nu $ distributions from the experimental data 
match those from the reweighted simulated data.
The effect of this reweighting on the extracted SDMEs is small
in most cases,
and the difference between final SDMEs and
those obtained with reweighted MC events is taken as systematic uncertainty.
\end{itemize}

The total systematic uncertainties are obtained by adding the above described components in quadrature. Table~\ref{tab1} gives the values for the total kinematic region.
The individual contributions i) - v) to the systematic uncertainty
for each SDME are compiled in Table \ref{atab1} in the Appendix.
When averaged over all SDMEs it appears that the group i) systematics dominates by contributing almost half of the systematic uncertainties, while about one-fifth contributions arise from both group ii) and group iv) systematics.
 In most cases, the statistical uncertainty is comparable to or smaller than the total systematic one.

\section{Results}
\label{Re}
\subsection{SDMEs for the total kinematic region}
\begin{figure}[hbt!]\centering
\includegraphics[width=12cm]{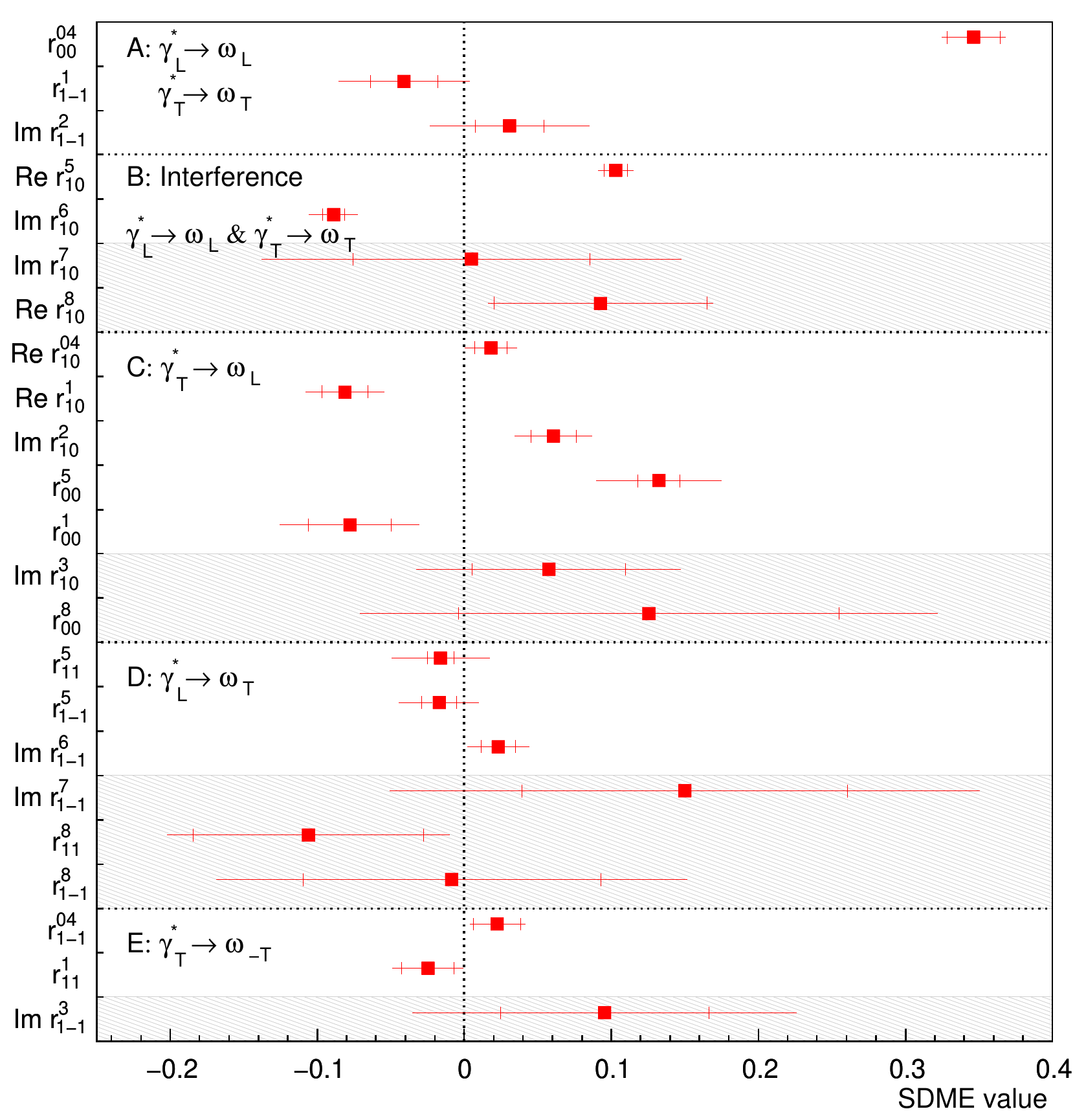}
\caption{
The 23 SDMEs for exclusive $\omega$ leptoproduction extracted in the total  COMPASS kinematic
region with $ \langle Q^2 \rangle = 2.13$~(GeV/$c$)$^2$, $\langle W\rangle =7.6$~GeV/$\it {c}^2$, $ \langle p^{2}_{\rm T}\rangle = 0.16$
(GeV/$c$)$^2$. 
Inner error bars represent statistical uncertainties and
outer ones statistical and systematic uncertainties added in
quadrature. Unpolarised (polarised) SDMEs are displayed in %the 
unshaded (shaded) areas.
}
\label{results}
\end{figure}

 The SDMEs extracted in the total kinematic region 
 1.0~(GeV/$c$)$^2$ $< Q^{2} < 10.0$~(GeV/$c$)$^2$, 5.0~GeV/$c^2$  $< W < 17.0$~GeV/$c^2$ and 
 0.01~(GeV/$c$)$^2$ $<p^{2}_{\rm T} < 0.5$~(GeV/$c$)$^2$,
with mean values
$\langle Q^{2} \rangle= 2.13$~(GeV/$c$)$^2$, $\langle W \rangle= 7.6$~GeV/$\it {c}^2$ and
$\langle p^{2}_{\rm T} \rangle = 0.16$~(GeV/$c$)$^2$ are presented in
Fig.~\ref{results} and %in 
Table~\ref{tab1}. These SDMEs are presented in five classes corresponding
to different helicity transitions. For the SDMEs in class A,
the dominant contributions are related to the squared amplitudes for 
 transitions from
longitudinal virtual photons to  longitudinal vector mesons, $\gamma^*_L \to
V_L$, and from transverse virtual photons to transverse vector mesons, $\gamma^*_T
\to V_T$. In class B, the dominant  terms correspond to the interference 
between amplitudes for the two aforementioned transitions.
 The main terms in most of the SDMEs for classes C, D and E
are proportional to the products of  small
amplitudes describing $\gamma^*_T \to V_L$, $\gamma^*_L \to V_T$ and $\gamma^*_T \to V_{-T}$ transitions respectively, and the large helicity-conserving amplitudes $T_{11}$ or $T_{00}$.\\
\indent In Fig.~\ref{results},  polarised SDMEs are shown in %the 
shaded areas.
The experimental uncertainties of these SDMEs are larger
than those of the unpolarised SDMEs because the lepton-beam
polarisation is smaller than unity ($|P_{b}| \approx 80\%$) and in the
expressions for the
angular distributions (see Eq.~(\ref{eqang3}))  
they are multiplied by the small kinematic factor
$|P_{b}|\sqrt{1- \epsilon} $,
where $ \epsilon \approx 0.96$.
In the calculation of the statistical uncertainty, the correlations between the various SDMEs are taken into account.

\begin{table}[hbt!] 
\begin{center}
\caption{\label{tab1} The 23 unpolarised and polarised SDMEs %values 
 for the total COMPASS kinematic region,
shown in the same order as in Fig.~\ref{results} for classes A to E.    
The first uncertainties are statistical, the second  systematic.}
\renewcommand{\arraystretch}{1.2}
\begin{tabular}{|c|r@{\,$\pm$\,}r@{\,$\pm$\,}r|}
\hline
 SDME &\multicolumn{3}{c|}{}     \\
\hline
$r^{04}_{00}$    &$  0.346$&$ 0.018$&$ 0.008$ \\
$r^1_{1-1}$      &$ -0.041$&$ 0.023$&$ 0.038$ \\
Im $r^2_{1-1}$   &$  0.031$&$ 0.023$&$ 0.049$ \\
\hline
Re $r^5_{10}$    &$  0.103$&$ 0.008$&$ 0.010$ \\
Im $r^6_{10}$    &$ -0.089$&$ 0.007$&$ 0.015$ \\
Im $r^7_{10}$    &$  0.005$&$ 0.081$&$ 0.118$ \\
Re $r^8_{10}$    &$  0.093$&$ 0.072$&$ 0.025$ \\
\hline
Re $r^{04}_{10}$ &$  0.018$&$ 0.011$&$ 0.014$ \\
Re $r^1_{10}$    &$ -0.081$&$ 0.016$&$ 0.022$ \\
Im $r^2_{10}$    &$  0.061$&$ 0.015$&$ 0.021$ \\
$r^5_{00}$       &$  0.132$&$ 0.014$&$ 0.039$ \\
$r^1_{00}$       &$ -0.078$&$ 0.028$&$ 0.040$ \\
Im $r^3_{10}$    &$  0.057$&$ 0.052$&$ 0.073$ \\
$r^8_{00}$       &$  0.125$&$ 0.130$&$ 0.148$ \\
\hline
$r^5_{11}$       &$ -0.016$&$ 0.009$&$ 0.032$ \\
$r^5_{1-1}$      &$ -0.017$&$ 0.012$&$ 0.025$ \\
Im $r^6_{1-1}$   &$  0.023$&$ 0.011$&$ 0.018$ \\
Im $r^7_{1-1}$   &$  0.150$&$ 0.111$&$ 0.168$ \\
$r^8_{11}$       &$ -0.106$&$ 0.078$&$ 0.056$ \\
$r^8_{1-1}$      &$ -0.009$&$ 0.101$&$ 0.124$ \\
\hline
$r^{04}_{1-1}$   &$  0.022$&$ 0.016$&$ 0.011$ \\
$r^1_{11}$       &$ -0.025$&$ 0.018$&$ 0.017$ \\
Im $r^3_{1-1}$   &$  0.095$&$ 0.071$&$ 0.110$ \\

\hline
\end{tabular}
\end{center}
\end{table}

\subsection{Dependences of SDMEs on $Q^{2}$, $p^{2}_{\rm{T}}$ and $ W $}

\begin{figure*}[hbt!]\centering
\includegraphics[width=15cm]{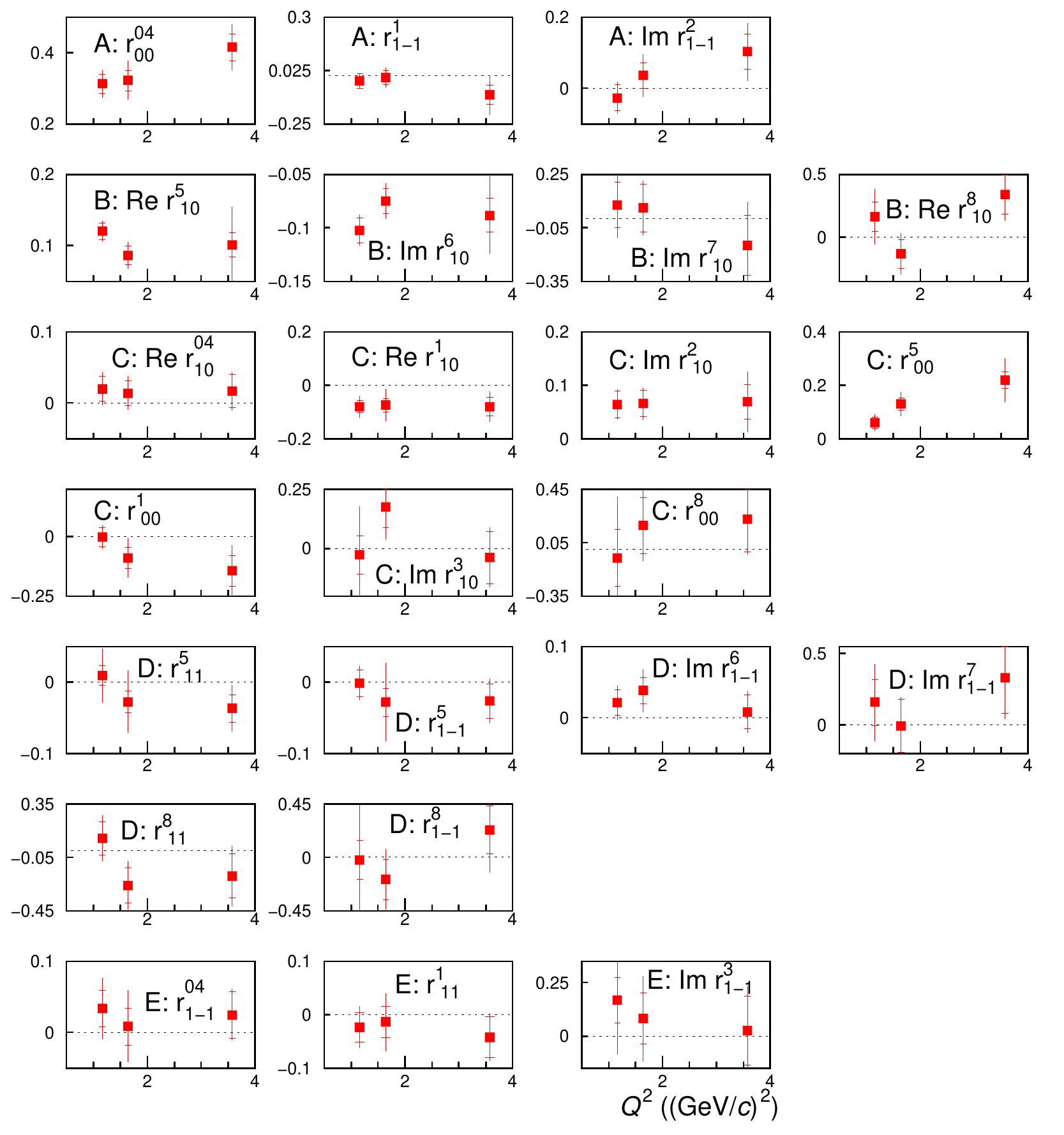}
\caption{ 
 $Q^2$ dependence of the measured 23 SDMEs. 
 The capital letters A to E denote the class, to which the SDME belongs. 
Inner error bars represent statistical uncertainties and 
outer
ones statistical and systematic uncertainties added in
quadrature.
}
\label{q1_testnsys}
\end{figure*}

\begin{figure*}[hbt!]\centering
\includegraphics[width=15cm]{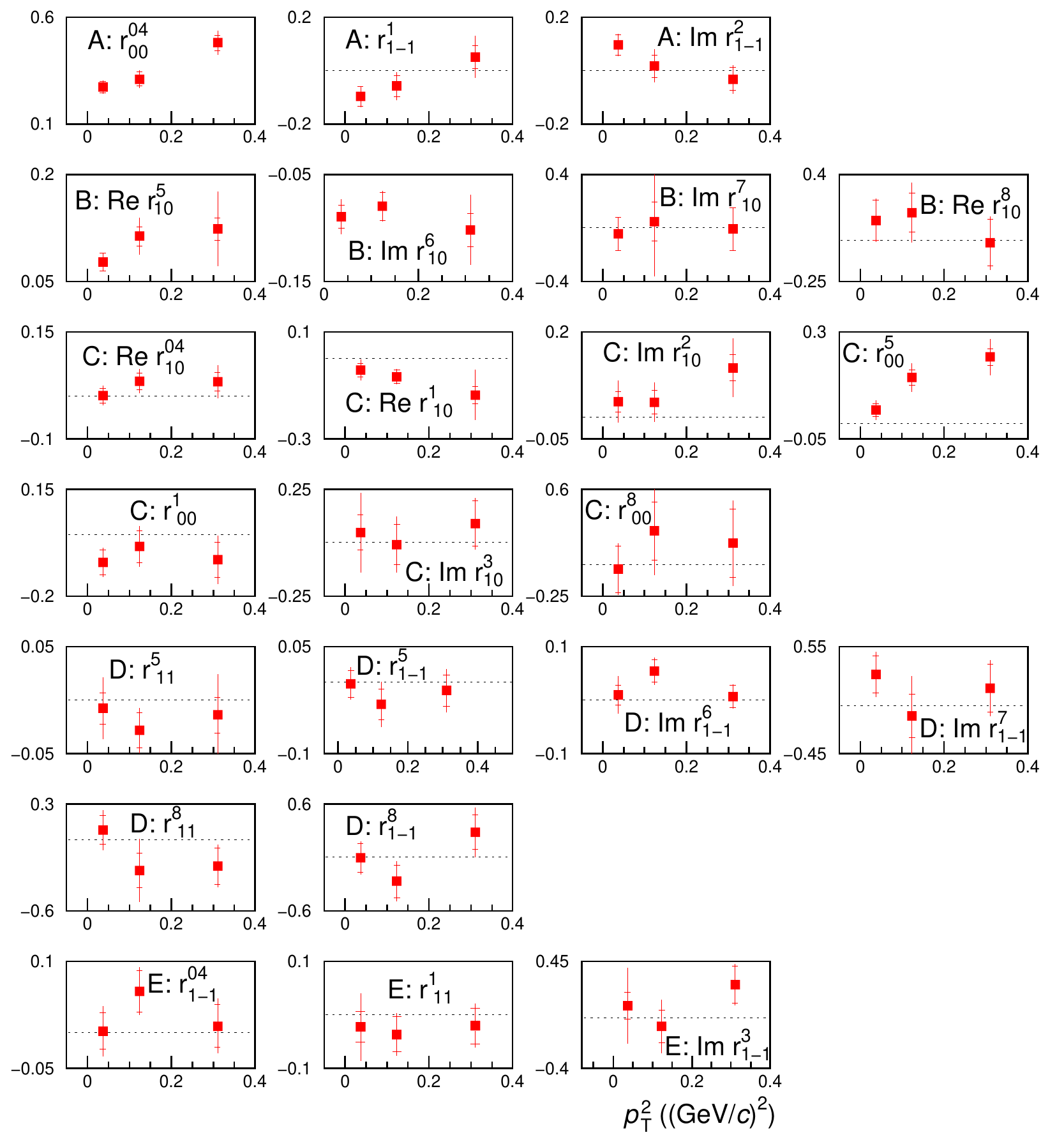}
\caption{ 
$p^{2}_{\rm{T}}$ dependence of the measured 23 SDMEs. 
The capital letters A to E denote the class, to which the SDME belongs.
%The SDMEs are divided into five classes A, B, C, D, and E defined  in Sec.~\ref{Re}.
Inner error bars represent statistical uncertainties and
outer ones statistical and systematic uncertainties added in
quadrature.
}
\label{pt1_testnsys}
\end{figure*}

\begin{figure*}[hbt!]\centering
\includegraphics[width=15cm]{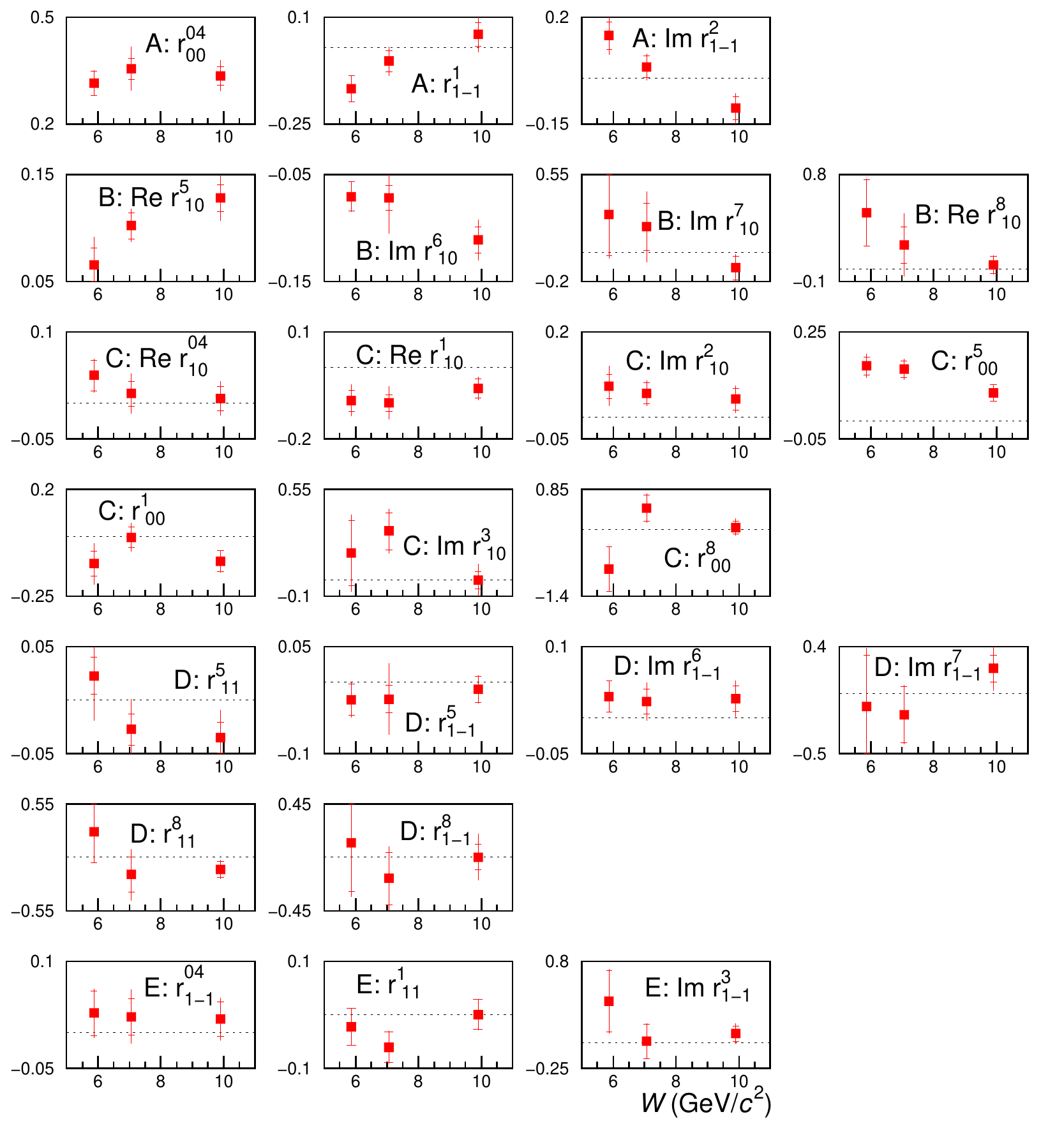}
\caption{ 
 $W$ dependence of the measured 23 SDMEs. 
 The capital letters A to E denote the class, to which the SDME belongs. 
% The SDMEs are divided into five classes A,  B, C, D, and  E defined  in Sec.~\ref{Re}.
Inner error bars represent statistical uncertainties and
outer ones statistical and systematic uncertainties added in
quadrature.
}
\label{w1_testnsys}
\end{figure*}

The kinematic dependences of the SDMEs on  $Q^{2}$, $p^{2}_{\rm{T}}$ and $ W $,
which have been determined in three bins for each of the variables, are shown in Figs.~\ref{q1_testnsys}, \ref{pt1_testnsys} and \ref{w1_testnsys}.
In Table~\ref{kinem-h}, the limits of the kinematic bins and the mean values of kinematic variables in the bins are given.
The values of SDMEs in bins of $Q^{2}$, $p^{2}_{\rm{T}}$ and
$ W $ are given in Table~\ref{tab2}, \ref{tab3} and \ref{tab4} respectively, in the Appendix.

\begin{table}

\begin{center}
\caption{ \label{kinem-h}
Kinematic binning and mean values for kinematic variables.} 

\renewcommand{\arraystretch}{1.2}
\begin{tabular}{|l@{\,}l@{\,<\,}c@{\,<\,}l@{\,}l|r@{\,}l|}
\hline
\multicolumn{5}{|c}{bin}&\multicolumn{2}{|c|}{$ \langle Q^{2} \rangle$}   \\
\hline
1.0\phantom{0}&(GeV/$c$)$^2$ & $Q^{2} $&\phantom{1}1.35& (GeV/$c$)$^2$  & 1.16&(GeV/$c$)$^2$ \\
1.35          &(GeV/$c$)$^2$ & $Q^{2} $&\phantom{1}2.05& (GeV/$c$)$^2$  & 1.64&(GeV/$c$)$^2$ \\
2.05          &(GeV/$c$)$^2$ & $Q^{2} $&10.0& (GeV/$c$)$^2$  & 3.61&(GeV/$c$)$^2$ \\
\hline
\multicolumn{5}{|c}{bin}&\multicolumn{2}{|c|}{$ \langle p^2_{\rm{T}}\rangle $}   \\
\hline
0.01          &(GeV/$c$)$^2$ &$p^{2}_{\rm{T}}$&\phantom{2}0.07&(GeV/$c$)$^2$ & 0.037&(GeV/$c$)$^2$\\
0.07          &(GeV/$c$)$^2$ &$p^{2}_{\rm{T}}$&\phantom{2}0.19&(GeV/$c$)$^2$ & 0.125&(GeV/$c$)$^2$\\
0.19          &(GeV/$c$)$^2$ &$p^{2}_{\rm{T}}$&\phantom{2}0.5 &(GeV/$c$)$^2$ & 0.310&(GeV/$c$)$^2$\\
\hline
\multicolumn{5}{|c}{bin}&\multicolumn{2}{|c|}{$ \langle W \rangle $}   \\
\hline
5.0           &GeV/$c^2$&$ W $&\phantom{2}6.4 & GeV/$c^2$ & 5.87&GeV/$c^2$\\
6.4           &GeV/$c^2$&$ W $&\phantom{2}7.9 & GeV/$c^2$ & 7.06&GeV/$c^2$ \\
7.9           &GeV/$c^2$&$ W $&17.0 & GeV/$c^2$ & 9.90&GeV/$c^2$ \\
\hline
\end{tabular}
\end{center}
\end{table}

\section{Discussion}

\subsection {Test of the SCHC hypothesis}
\indent   In case of SCHC, only the seven SDMEs of classes A and B are not
restricted to vanish, while all SDMEs from classes C, D, and E should be equal to
zero. Six of the SDMEs in classes A and B have to fulfil the following
relations
\cite{Schill}:\
~~~~~~~~~~~~~~~~~~~~~~~~\\
\begin{eqnarray}
r_{1-1}^1 &=&-\mathrm{Im} \{r_{1-1}^2\},\nonumber\\
\mathrm{Re}\{r_{10}^5\} &=&-\mathrm{Im}\{r_{10}^{6}\},\nonumber\\
\mathrm{Im} \{r_{10}^{7}\} &=&~~\: \mathrm{Re} \{r_{10}^{8}\}.\nonumber
\end{eqnarray}
Within uncertainties, the extracted SDMEs are consistent with these
relations:
\begin{eqnarray}
r^{1}_{1-1}+\mathrm{Im}\{r^{2}_{1-1}\}&=& -0.010 \pm 0.032  \pm
0.047,\nonumber\\
 \mathrm{Re}\{r^{5}_{10}\}+  \mathrm{Im}\{r^{6}_{10}\}&=&~~\, 0.014 \pm
 0.011 \pm 0.013 ,\nonumber\\
 \mathrm{Im}\{r^{7}_{10}\} -\mathrm{Re}\{r^{8}_{10}\} &=&-0.088  \pm  0.110
 \pm
 0.196.  \nonumber
\end{eqnarray}

However, for the transitions $\gamma^*_{\rm T} \to V_{\rm L}$ of class C the non-zero values of SDMEs $r^5_{00}$ 
and $\mathrm{Re}\{r^1_{10}\}$ show SCHC violation at the level of 
three standard deviations of the statistical uncertainty.
In the GK model
\cite{Goloskokov:2009},
these SDMEs are related to  the chiral-odd GPDs 
$H_{\rm T}$ and ${\bar E}_{\rm T}$ coupled to the higher-twist wave function
of the meson. The kinematic dependences of these  SDMEs, as presented in
Section \ref{Re}, may
help to further constrain  the model.

 %}

\subsection{UPE contribution in exclusive $\omega$ meson production}

The existence of UPE transitions in 
exclusive   $\omega$ production 
 can  be tested by examining linear combinations of SDMEs
 %\magenta{[WDN 6/12 13h30: I propose to simply omit the text that follows up the the formula, as it's repeated anyway in a correct way just after.]} , which \red{'only' not precise because NPE enter in denominator of Eq. (30)} contain UPE amplitudes, 
such as
\begin{equation}
u_1=1-r^{04}_{00}+2r^{04}_{1-1}-2r^{1}_{11}-2r^{1}_{1-1}.
\label{uu1}   
\end{equation}
The quantity $u_1$ can be expressed in terms of helicity amplitudes  as

\begin{equation}
u_1=\widetilde{\sum}\frac{4\epsilon|U_{10}|^2+2|U_{11}+U_{-11}|^2}{\mathcal{N}}.
\label{u1u}   
\end{equation}
Since the numerator depends only on UPE amplitudes, 
a $u_1$ value different from zero indicates non-zero contribution from UPE transitions.
For the total kinematic
region of COMPASS $u_1$ is equal to  0.830 $\pm$ 0.073 $\pm$ 0.049, which is a clear signal
of a large UPE contribution. Additional information on UPE amplitudes
is obtained from the SDME combinations   
\begin{equation}
u_2=r^{5}_{11}+r^{5}_{1-1}
\label{uu2}
\end{equation} 
and
\begin{equation}
u_3=r^{8}_{11}+r^{8}_{1-1},
\label{uu3}   
\end{equation}
which in terms of
helicity amplitudes can be combined into
\begin{equation}
u_2 + iu_3
=\sqrt2\widetilde{\sum}\frac{(U_{11}+U_{-11})U^{*}_{10}}{\mathcal{N}}.
\label{u2u3n}
\end{equation}
For COMPASS, $u_2$ = $-$0.033 $\pm$ 0.016 $\pm$ 0.043 and $u_3$ = $-$0.114 $\pm$ 0.126 $\pm$ 0.099 are obtained, which are consistent with zero at the present accuracy of the data.

\begin{figure*}[hbt!]\centering
\includegraphics[width=15cm]{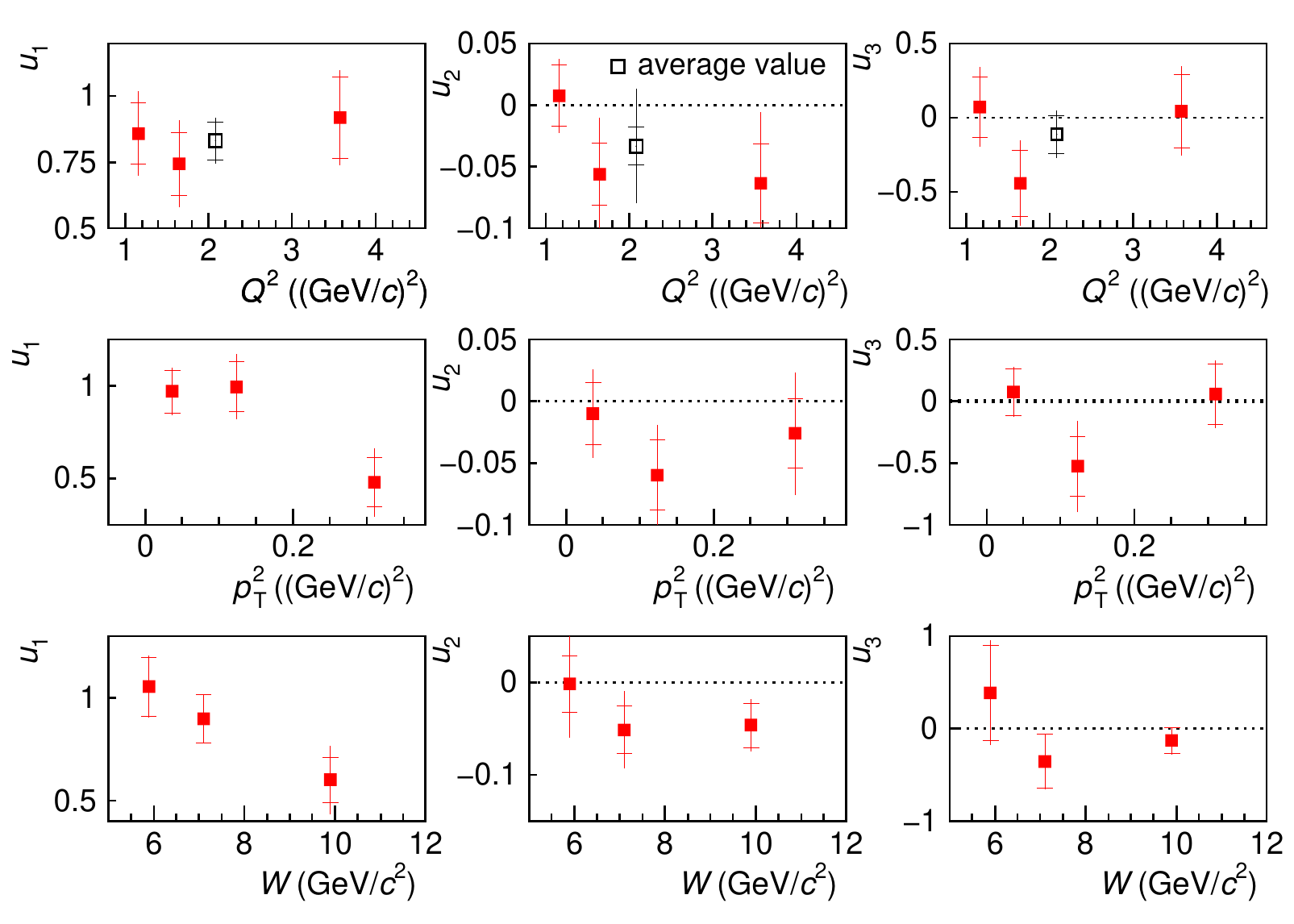}
\caption{
$Q^{2}$, $p^{2}_{\rm{T}}$, and $W$ dependences of $u_{1}$, $u_{2}$, $u_{3}$.
The open symbols represent the values over the total kinematic region. 
Inner error bars represent statistical uncertainties and
outer ones statistical and systematic uncertainties added in
quadrature.}
\label{depq2ptw}
\end{figure*}

In  Fig.~\ref{depq2ptw} the dependence of the quantities  $u_{1}$, $u_{2}$, $u_{3}$
 on $Q^{2}$, $p^{2}_{\rm {T}}$, and $W$ is  presented.
 The quantity  
 $u_{1}$ decreases with increasing $W$ and $p^{2}_{\rm{T}}$,  which indicates that the UPE contribution becomes 
 smaller, while $u_{2}$, $u_{3}$ fluctuate around zero.

More detailed information on the $W$dependence of certain UPE transitions  in terms of helicity amplitudes can be obtained by considering the difference of the following two class-A SDMEs~\cite{DC-24}: 
\begin{eqnarray}
r_{1-1}^1  = \frac {1}{2\mathcal{N}}\widetilde{\sum}\{|T_{11}|^2+|T_{1-1}|^2 \nonumber \\
-|U_{11}|^2-|U_{1-1}|^2\},
\label{sdprz}   
\end{eqnarray}
\begin{eqnarray}
\mathrm{Im} \{r_{1-1}^2 \}
=\frac {1}{2\mathcal{N}} \widetilde{\sum} \{-|T_{11}|^2+|T_{1-1}|^2
\nonumber\\  
+|U_{11}|^2 -|U_{1-1}|^2\},
\label{sdprz1}
\end{eqnarray}
which reads:
\begin{equation}
\mathrm{Im} \{r_{1-1}^2\} - r_{1-1}^1 = \frac
{1}{\mathcal{N}}\widetilde{\sum}(-|T_{11}|^2+|U_{11}|^2).
\label{difff} 
\end{equation}
For the total kinematic region, both SDMEs and their difference are close to zero. For the present data,   $\mathrm{Im} \{r^2_{1-1}\}-r^1_{1-1}$= 0.07$\pm$0.07 is obtained, hence $\widetilde{\sum}|U_{11}|^2 \approx \widetilde{\sum}|T_{11}|^2$. 
By applying Eq.~(\ref{sum-two}), Eq.~(\ref{difff}) can be rewritten as follows:    
\begin{eqnarray}
\mathrm{Im}
\{r^2_{1-1}\}-r^1_{1-1}
=\frac{1}{\mathcal{N}}
(-|T_{1 \frac{1}{2}1 \frac{1}{2}}|^2-|T_{1 -\frac{1}{2}1 \frac{1}{2}}|^2
\nonumber \\
+|U_{1 \frac{1}{2}1 \frac{1}{2}}|^2+|U_{1 -\frac{1}{2}1\frac{1}{2}}|^2).
\label{appl}  
\end{eqnarray}
Bilinear contributions
of nucleon helicity-flip amplitudes  are suppressed by
a factor $(\sqrt{-t^{\prime}}/M)^2$, 
where $t'$ is a measure of the transverse momentum of the vector meson  
with respect to the direction of the virtual photon. 
Neglecting these bilinear contributions yields: 
\begin{equation}
\mathrm{Im}
\{r_{1-1}^2\} - r_{1-1}^1 \approx \frac{1}{\mathcal{N}} (|U_{11}|^2 - |T_{11}|^2). 
\end{equation}
 %as  W   increases

 %Beside of SDMES for the total kinematic region, the kinematic dependences

\begin{table*}
\renewcommand{\arraystretch}{1.2}
\caption{\label{wdepen} $W$ dependence of SDMEs  $r^{1}_{1-1}$, Im $r^{2}_{1-1}$
and their difference}
\begin{center}
\begin{tabular}{|c|l|l|l|}
\hline
  $\langle W \rangle$ (GeV/$c^2$) &\multicolumn{1}{c|}{5.9} & \multicolumn{1}{c|}{7.1}  &\multicolumn{1}{c|}{9.9}\\
\hline
 $r^{1}_{1-1}$             &$-0.134          \pm0.043\pm0.32$  &$ -0.044          \pm0.036\pm0.33$  &$\phantom{-} 0.052\pm0.038\pm0.047$\\
\hline
Im $r^{2}_{1-1}$            &$\phantom{-}0.139\pm0.044\pm0.46$  &$\phantom{-}0.037 \pm0.036\pm0.24$  &$-0.098           \pm0.038\pm0.033$\\
\hline
Im $r^{2}_{1-1}-r^{1}_{1-1}$&$\phantom{-}0.273\pm0.061\pm0.046$ &$\phantom{-}0.081\pm0.050\pm0.041$ &$-0.151\pm0.053    \pm0.057$\\
\hline
\end{tabular}
\end{center}
\end{table*}
In  Table~\ref{wdepen}, the values of the  SDMEs
 $ r^{1}_{1-1}$ and Im$\{{r^{2}_{1-1}}\}$ and their difference are shown   as a function of $\langle W \rangle$.
The difference is large and positive at
$ \langle W \rangle = 5.9$~GeV/$c^2$, i.e.\ $ |U_{11}|>|T_{11}|$. 
For $ \langle W \rangle = 7.1$~GeV/$c^2$, $ |U_{11}| \approx |T_{11}|$ holds and
for $ \langle W \rangle = 9.9$~GeV/$c^2$ the situation is reversed:  $ |U_{11}| < |T_{11}|$.\\ 

A substantial contribution of UPE transitions
in hard exclusive $\omega$ meson electroproduction was observed at HERMES~\cite{HERMES:2014}. In their total
 kinematic range% of HERMES
 , with mean values $\langle Q^{2} \rangle=
2.4$~(GeV/$c$)$^2$, $ \langle W
\rangle = 4.8$~GeV/$c^2$ and $ \langle t' \rangle = 0.08$
(GeV/$c$)$^2$, for the proton target they found 
$u_1 = 1.1 \pm 0.09 \pm 0.12$ and %the SDMEs difference  
 $\mathrm{Im} \{r^2_{1-1}\}-r^1_{1-1} = 0.35\pm0.04\pm0.05$. 
The latter value indicates that  %the 
$|U_{11}|^2
> |T_{11}|^2$ %for HERMES 
in their kinematic range. %Their HERMES also demonstrates 
Also they observed that the quantity $u_1$, when averaged over the total range of $W$,
%the quantity $u_1$ 
increases (decreases) with increasing values of $Q^2$ ($t'$).

A quantitative comparison of COMPASS and HERMES results is not
straightforward, because the covered kinematic regions %are different. Although they 
only partially overlap, and COMPASS covers
significantly wider ranges of $W$ and $p^2_{\rm T}$. %Besides, WEN 
It is important to note here that, when studying the kinematic dependences of
the measured observables, %the values 
results are extracted
in one-dimensional intervals of a given kinematic variable, while averaging
over the full ranges of the other two %other 
variables. %These ranges are different in each 
As the two experiments have only partially overlapping kinematic ranges, %, which affects 
the results %of 
after averaging cannot be directly compared.

When neglecting the observed $Q^2$ and $t'$ ($p^2_{\rm T}$) dependences, which 
exhibit opposite trends, one can compare the HERMES %estimate of 
result on $u_1$ for 
their total kinematic range to 
the COMPASS result shown at the lowest $W$ value in 
Fig.~\ref{depq2ptw}. Similarly, %HERMES 
HERMES result on
$\mathrm{Im} \{r^2_{1-1}\}-r^1_{1-1}$ can be compared to the corresponding value from
COMPASS at $ \langle W
\rangle = 5.9$~GeV/$c^2$%that 
, which is shown in Table~\ref{wdepen}.
Within uncertainties the results %for both observables 
from the two experiments are consistent for both observables.

Altogether, the main COMPASS results presented in this subsection, i.e.\ the $W$ dependence of
$u_1$ as well as that of %the SDMEs difference 
$\mathrm{Im} \{r^2_{1-1}\}-r^1_{1-1}$ 
indicate that
the UPE contribution
decreases with increasing $W$ without vanishing towards largest $W$ values
accessible at COMPASS. In the GK model, UPE is described by the GPDs $\widetilde{H}^{f}$ and $\widetilde{E}^{f}$ (non-pole), and by the pion-pole  contribution treated as a one-boson
exchange~\cite{GK:epjA-2014}. The latter one, which is a sizeable contribution, results in
a significantly faster decrease of the predicted UPE contribution with increasing $W$ 
than that measured
at COMPASS.

 %\red{ [AS] I do not think that we can conclude like following text; NPE %in fact grows that is consistent with growth of exclusive omega %production cross section with increasing W as observed at ZEUS; I %propose to stay with previous formulation of conclusions (starting from %"It is also consistent..."}  
 %\magenta{A possible interpretation is that the NPE amplitude $T_{11}$ 
 %shows no strong $W$ dependence, while the UPE amplitude $U_{11}$, 
 %which is mostly fed by pion-pole exchange, decreases but not vanishes %towards the high-$W$ limit of the kinematic range.}  
~~~~~~\\
\subsection{The NPE-to-UPE asymmetry of the transverse cross section for the transition $\gamma^*_T \rightarrow V_T$} 

Another observable that is sensitive to the relative contributions of UPE and
NPE amplitudes is the 
NPE-to-UPE asymmetry of the transverse differential cross section for the transition $\gamma^*_T \rightarrow V_T$. It is
defined~\cite{GK:epjA-2014} as\footnote{In Ref.~\cite{HERMES:2014} 
a different
definition of the asymmetry is used.}
\begin{eqnarray}
\nonumber
P=\frac{d\sigma^N_T(\gamma^*_T \rightarrow V_T) 
- d\sigma^U_T  (\gamma^*_T \rightarrow V_T)}
{d\sigma^N_T (\gamma^*_T \rightarrow V_T) +
d\sigma^U_T  (\gamma^*_T \rightarrow V_T)} \\
=\frac{2 r^1_{1-1}}{1-r^{04}_{00}-2 r^{04}_{1-1}},
\label{asymmGK}
\end{eqnarray}
where the superscript %s 
$N$ and $U$  
denotes the part of the cross section that is
fed by NPE and UPE transitions,
respectively.

\begin{figure*}
 %[hbtc!]
 \centering
 \includegraphics[width=16cm]{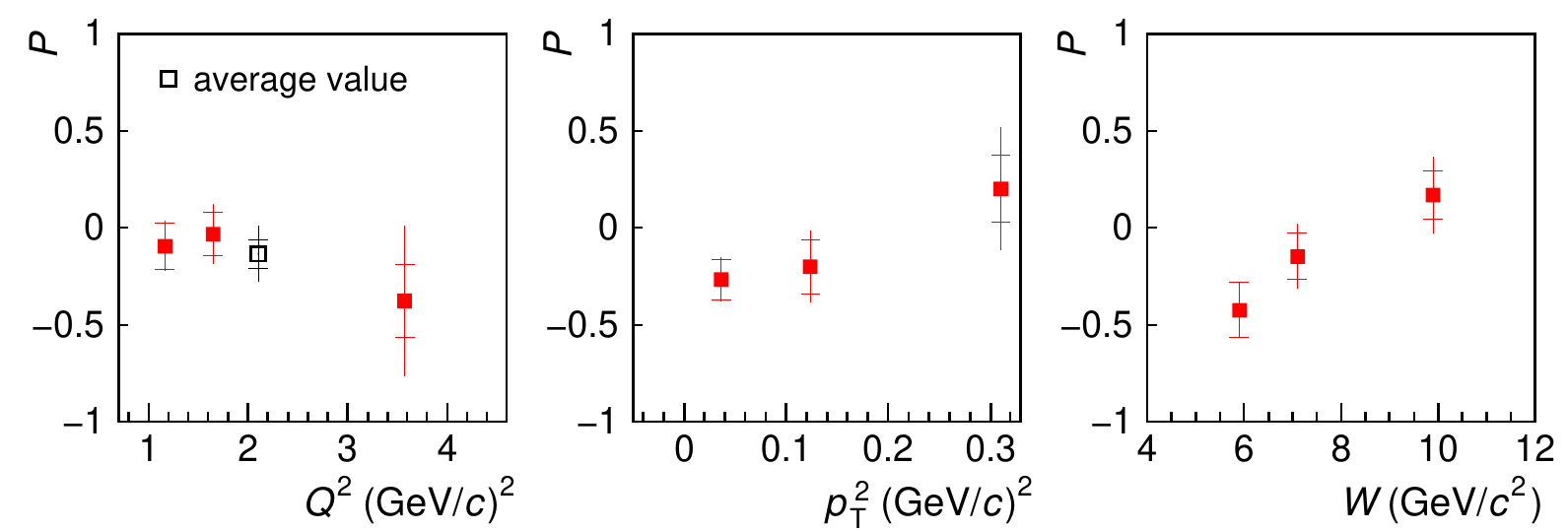}
 \caption{ $Q^{2}$, $p^{2}_{\rm{T}}$ and $W$ dependences of the NPE-to-UPE
 asymmetry of  
 the transverse cross section for the transition $\gamma^*_T \rightarrow V_T$.
 The open symbol represents the value over the total kinematic region.
 Inner error bars represent statistical uncertainties and outer
 ones statistical and systematic uncertainties added in
 quadrature. }
 \label{asyupenpenew}
 \end{figure*}

The value of $P$ obtained in the total kinematic region is $-0.007
\pm 0.076 \pm 0.125$, which indicates that the UPE and NPE
contributions averaged over the whole kinematic range of COMPASS are of similar size.
The kinematic dependences of the asymmetry $P$ are shown in Fig.~\ref{asyupenpenew}. The UPE contribution
dominates at small values of $W$ and $p^2_{\rm T}$
and decreases with increasing values of these kinematic variables.
At large values of
$W$ and $p^2_{\rm T}$, the NPE contribution becomes dominant, %still with 
while a non-negligible UPE contribution remains.
No significant $Q^2$ dependence 
of the asymmetry is observed.

\subsection{Longitudinal-to-transverse cross-section ratio}

In order to evaluate the longitudinal-to-transverse virtual-photon  differential
cross-section ratio
\begin{equation}
R= \frac{d\sigma_{L}(\gamma^{*}_{L} \to V)}{ d\sigma_{T}(\gamma^{*}_{T} \to
V)},
\end{equation}
the quantity $R'$ can be used:
\begin{equation}
R'  =
\frac{1}{\epsilon}\frac{r^{04}_{00}}{1-r^{04}_{00}}.
\label{sigto} 
\end{equation}
Using expressions defining $r^{04}_{00}$ and $1-r^{04}_{00}$
in terms of helicity amplitudes~\cite{Schill,DC-24},
one obtains
\begin{eqnarray}
R'= 
\frac{1}{\epsilon} 
\dfrac{
\widetilde{\sum}(\epsilon|T_{00}|^2+|T_{01}|^2+|U_{01}|^2)}
%/
% ~~~~~~~~~~~~~~~~~~~~~~~~~~
% \nonumber \\
{
\widetilde{\sum}\{|T_{11}|^2+|U_{11}|^2+|T_{1-1}|^2+|U_{1-1}|^2 
% ~~~~~~~~~~~~~
% \nonumber \\
+2\epsilon (|T_{10}|^2+|U_{10}|^2)\}}.
%~~~~~~~~~~~~
\label{Rprimhelamp}
\end{eqnarray}
The quantity $R'$ may be interpreted as the 
longitudinal-to-transverse
ratio of ``effective''
cross sections for the production of vector mesons that are polarised
longitudinally or 
transversely irrespective of the virtual-photon polarisation. In case of 
SCHC, $R'$ is equal to $R$. 
In spite of the observed clear violation of SCHC at COMPASS, we use the approximate relation 
$R \approx R'$. % is used. 
The accuracy of this approximation 
is estimated
%~\cite{lastGK} 
using
the GK model~\cite{GK:epjC-2014,GK:epjA-2014} and
the resulting uni-directional systematic uncertainty is
found to be about $+15\%$ on average, while its magnitude ranges
between 3\% and 47\% with increasing $W$ and between 6\% and 28\% with increasing $p_T^2$.

\begin{figure*}
\centering
\includegraphics[width=16cm]{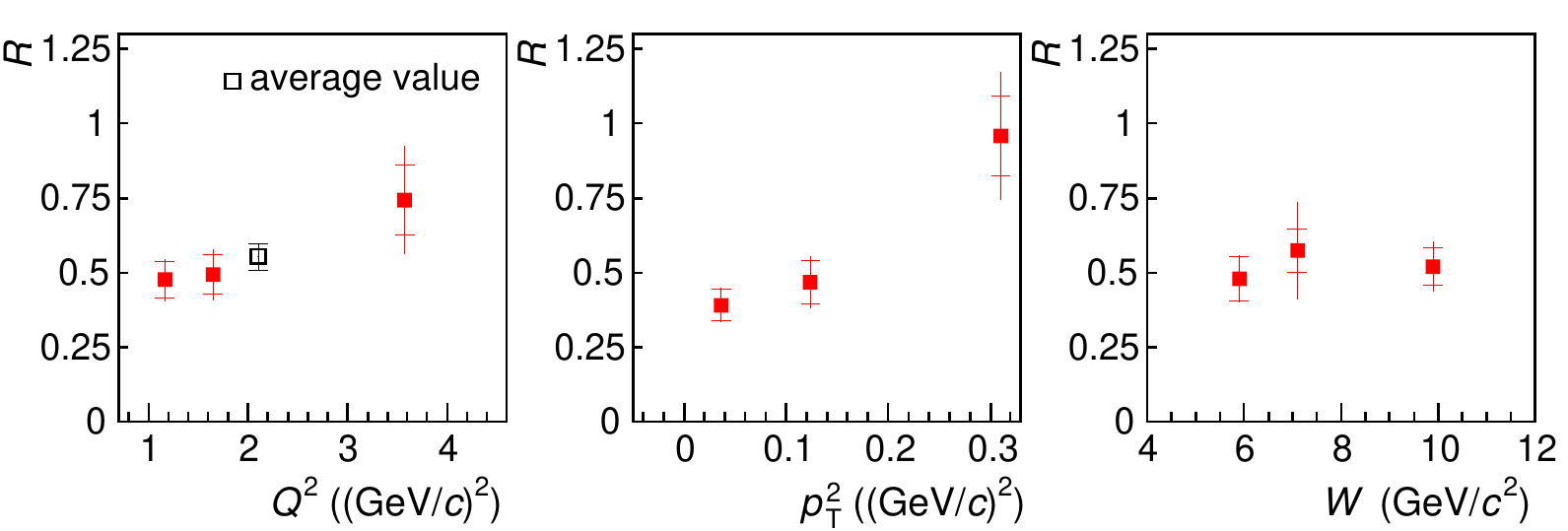}
\caption{ $Q^{2}$, $p^{2}_{\rm{T}}$ and $W$ dependences of the longitudinal-to-transverse cross-section ratio $R$. The open symbol represents the value %over 
obtained for the total
kinematic region.
Inner error bars represent statistical uncertainties and
outer ones statistical and systematic uncertainties added in
quadrature. 
Note that the additional positive uni-directional systematic uncertainty due to the approximation 
$R \approx R'$ is not shown here, see text for details.}
\label{depq2ptwa}
\end{figure*}

For the total kinematic region, the ratio $R$ is found to
be $0.553 \pm 0.044_{\rm stat} \pm 0.020_{\rm syst} \ _{- \ 0}^{+ \ 0.082}\rvert_{\rm appr}$.
Here, the third uncertainty is 
the systematic one
due to the approximation  $R \approx R'$.
The kinematic dependences of $R$ are  shown in Fig.~\ref{depq2ptwa}.
The ratio appears to increase as $Q^{2}$ and $p^{2}_{\rm{T}}$ increase, which indicates an increase of the fraction of longitudinally polarised vector mesons, while it shows 
no significant change over the $W$ range.

~~~~~~\\
 %\begin{figure*}
 %\centering
 %\includegraphics[width=20cm]{Fig/asyupenpeNew_pub.pdf}
 %\caption{
 % {\color{red} New Caption to be written by Andrzej }}
 %\label{asyupenpeNew}
 %\end{figure*}

\subsection{Phase difference between amplitudes}

Using  Eq.~(\ref{deu}), the phase difference between the UPE amplitudes
$U_{11}$ and $U_{10}$ can be calculated \cite{HERMES:2014}:

\begin{equation}
\tan \delta_{U} =u_3/u_2 =
\frac{r^{8}_{11}+r^{8}_{1-1}}{r^{5}_{11}+r^{5}_{1-1}}.
\label{deu}
\end{equation}
 The phase difference %obtained 
 $\delta_{U}$
 for the total kinematic region is found to be
 $\delta_{U} = (-106.1 \pm 53.6 \pm 2.5)$ degrees.

 The absolute value of the phase difference $\delta_{N}$ between the NPE amplitudes
 $T_{11}$ and  $T_{00}$ can be calculated using  Eq.~(\ref{eq:cosdelta})
 from  Ref.~\cite{DC-24}:

\begin{equation}
\cos \delta_{N}= \frac{ 2  \sqrt{\epsilon}
(\mathrm{Re}\{r^{5}_{10}\}-\mathrm{Im}\{r^{6}_{10} \})}
{\sqrt{ r^{04}_{00}(1-r^{04}_{00}+r^{1}_{1-1}-\mathrm{Im}\{r^{2}_{1-1}\})}}.
\label{eq:cosdelta}
\end{equation}

The phase difference $\delta_{N}$ for the total kinematic region is found to be
 |$\delta_{N}$| = (33.1 $\pm$ 4.9 $\pm$ 7.2) degrees.
 
Using the polarised 
SDMEs, also the sign of $\delta_{N}$  can in principle be
determined %from 
using the following equation from  Ref.~\cite{DC-24}:

\begin{equation}
\sin \delta_{N}= \frac{ 2  \sqrt{\epsilon}
(\mathrm{Re}\{r^{8}_{10}\}+\mathrm{Im}\{r^{7}_{10}\} ) }
{\sqrt{ r^{04}_{00}(1-r^{04}_{00}+r^{1}_{1-1}-\mathrm{Im}\{r^{2}_{1-1}\})}}.
\label{eq:sindelta}
\end{equation}

However, the large  experimental
uncertainties of the polarised 
SDMEs make this presently impossible. 

\subsection{Comparison with predictions of the GK model}
\begin{figure*}[hbt!]\centering
\includegraphics[width=12cm]{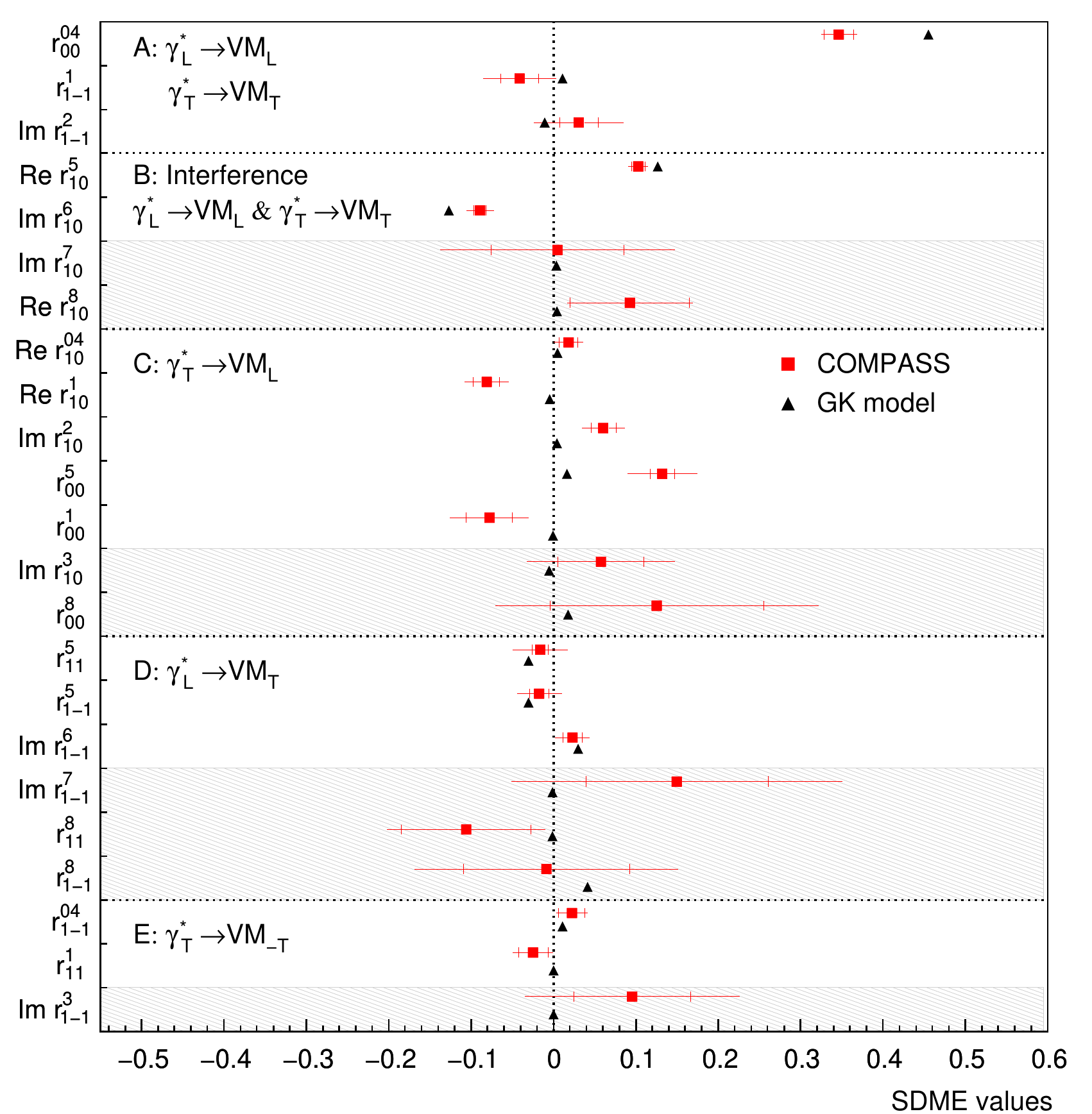}
\caption{\small{
Comparison of the measured SDMEs with  calculations of
the GPD model of
Goloskokov and  Kroll~\cite{GK:epjA-2014}. The calculations are obtained for
 $Q^{2}= 2.0$~(GeV/$c$)$^2$, $W=7.5$~GeV/$\it {c}^2$ and $p^{2}_{\rm T}=0.14$~(GeV$/c$)$^2$.
Inner error bars represent statistical uncertainties and
outer ones statistical and systematic uncertainties added in
quadrature.
}}
\label{results3}
\end{figure*}

In Fig.~\ref{results3} the 23 SDMEs for exclusive $\omega$ production, %s 
extracted in the total kinematic region of COMPASS, 
are compared with the predictions of the GPD model of Goloskokov and Kroll~\cite{GK:epjC-2014, GK:epjA-2014} for hard exclusive vector-meson leptoproduction.
In this version of the model, contributions from %the 
chiral-odd GPDs 
as well as from pion-pole exchange are included. The model was tuned
to HERMES results on SDMEs and spin asymmetries for exclusive $\rho ^0$ and $\omega$ production, which led to a satisfactory agreement between the model and the data. 

The predictions of the model shown in Fig.~\ref{results3} were obtained for exclusive $\omega$ production at
 $Q^{2}= 2.0$~(GeV/$c$)$^2$, $W=7.5$~GeV/$\it {c}^2$ and $p^{2}_{\rm T}=0.14$~(GeV$/c$)$^2$, close to the corresponding average kinematic values for COMPASS. 
 In the following, we concentrate on the most pronounced differences between %the 
 model predictions and
 experimental results.
 
The most noticeable differences are as follows: i) 
 the predicted value of SDME $r_{00}^{04}$, which represents the fraction of longitudinally polarised me\-sons in 
 the produced sample, is significantly larger than the measured one;
ii) the SDMEs dominated by the transitions  $\gamma^*_T \to \omega _L$
 (class C) %. In the model they 
 are in general close to zero in the model, while in the data several of them ($r^5_{00}$, $\mathrm{Re} \{r^1_{10}\}$) indicate a clear violation of the SCHC hypothesis. 
 
 A characteristic prediction of the model is a strong decrease of the UPE contribution with increasing values of $W$. The predicted values of the quantity $u_1$ at $\langle Q^{2} \rangle= 2.13$~(GeV/$c$)$^2$ 
and $\langle p^{2}_{\rm T} \rangle = 0.16$~(GeV/$c$)$^2$ are
 equal to 1.01, 0.39 and 0.07 for $W$ values of 5.0~GeV/$c^2$,
7.1~GeV/$c^2$ and 11.0~GeV/$c^2$
 respectively.  
A comparison of these predictions 
to the results shown in the lower-left panel of 
Fig.~\ref{depq2ptw} 
shows that, while at the smallest accessible value of $W$ the prediction is consistent with the measured $u_1$, at  
large values of $W$ the model predicts a much stronger decrease with increasing 
$W$ 
and hence underestimates significantly the measured UPE contribution.

%\begin{figure*}\centering

\section{Summary}
Using exclusive  $\omega$ meson muoproduction on protons, we have
measured 23 
Spin Density Matrix Elements at 
the average COMPASS kinematics,
$\langle Q^{2} \rangle=$ 2.1~(GeV/$c$)$^2$, $\langle W \rangle = 7.6$~GeV/$c^2$
and $\langle p^{2}_{\rm T} \rangle = 0.16$~(GeV/$c$)$^2$.
The SDMEs are extracted in the kinematic region 
 $1.0~({\rm GeV}/c)^2 < Q^{2} < 10.0$ $({\rm GeV}/c)^2$, 5.0~GeV/$c^2< W <17.0$~GeV/$c^2$ and 0.01~(GeV/$c$)$^2$ $<p^{2}_{\rm T} < 0.5$~(GeV/$c$)$^2$,
which allows us to study %ed studies of 
their $Q^2$, $p^2_{\rm T}$ and $W$ dependences. 

Several SDMEs that are dominated by amplitudes %for 
describing %transitions 
$\gamma^*_{\rm T} \to \omega_{\,\rm L}$ transitions
indicate a considerable violation of the SCHC hypothesis. These SDMEs are expected to be sensitive to the chiral-odd GPDs 
$H_{\rm T}$ and ${\bar E}_{\rm T}$, which are coupled to the higher-twist wave function
of the meson. A particularly prominent effect is observed for
the SDME $r^5_{00}$, which strongly increases with increasing $Q^2$ and $p^2_{\rm T}$, and decreases
with increasing $W$.

Using %the 
specific observables that are constructed to be sensitive to contributions from transitions with unnatural-parity exchanges such as $u_1$, %quantity, the SDMEs difference  
$\mathrm{Im} \{r^2_{1-1}\}-r^1_{1-1}$ and the UPE-to-NPE asymmetry for the transverse cross section,
a strong $W$ dependence of the UPE contribution is observed.  %from the transitions with unnatural parity exchanges. 
At low values of $W$, we confirm the earlier observation by %from 
HERMES that
the amplitude of the UPE transition $\gamma^*_{\rm T} \to \omega_{\rm T}$ is larger than the
NPE amplitude for the same transition, i.e.\ $ |U_{11}|>|T_{11}|$. With increasing $W$ the UPE
contribution decreases and $ |U_{11}|<|T_{11}|$ at large $W$, still with a non-negligible UPE
contribution at the largest $W$ values accessible at COMPASS.

Altogether, the COMPASS results presented in this paper cover a
kinematic range that extends considerably beyond the ranges of earlier
experimental data on SDMEs for exclusive $\omega$ leptoproduction. They
provide important input for modelling GPDs, in particular they may
help to better constrain the amplitudes for UPE transitions and assess
the role of chiral-odd GPDs  in exclusive $\omega$ leptoproduction.

\section*{Acknowledgements}
We are indebted to Sergey Goloskokov and Peter Kroll for numerous fruitful discussions on the interpretation of
our results and for providing us with predictions of their model. We gratefully acknowledge the support of CERN management
and staff and the skill and effort of the technicians of our
collaborating institutions.
We acknowledge the support by the European Union’s Horizon
2020 research and innovation programme under grant agreement STRONG –
2020 - No 824093.

\clearpage
%\onecolumn
\section*{Appendix}
\setcounter{table}{0}
\renewcommand{\thetable}{A.\arabic{table}}

%Tables \ref{atab1}, \ref{tab2}, \ref{tab3}, and %\ref{tab4} with kinematic dependences of 23 SDMEs.
Table \ref{atab1} gives the various contributions to the systematic uncertainty of the 23 SDMEs and
Tables \ref{tab2}, \ref{tab3}, and \ref{tab4} list their kinematic dependences.

\begin{table*}[hbt!]
\renewcommand{\arraystretch}{1.2}
\begin{center}
\caption{\label{atab1} Uncertainties for each SDME value: in column 3 the statistical
uncertainty (``stat.''), in columns 4--8 the individual contributions for
each source of systematic uncertainty as defined in Sec. \ref{sec:sys}, in column
9 the total systematic uncertainty (``tot. sys.''), and in column 10 the
total uncertainty (``tot.'').}
\begin{tabular}{|
>{\centering}m{0.08\textwidth}|
r@{}l|
>{\centering}m{0.05\textwidth}|
r@{}l|
r@{}l|
r@{}l|
r@{}l|
r@{}l|
>{\centering}m{0.05\textwidth}|
>{\centering\arraybackslash}m{0.05\textwidth}|
}
\hline
 SDME &
 \multicolumn{2}{m{0.06\textwidth}|}{value}  & 
 stat. &
 \multicolumn{2}{>{\centering}m{0.07\textwidth}|}{beam charge} &
 \multicolumn{2}{>{\centering}m{0.08\textwidth}|}{$E_{\texttt{miss}}$} &
 \multicolumn{2}{>{\centering}m{0.08\textwidth}|}{back\-ground} &
 \multicolumn{2}{>{\centering}m{0.07\textwidth}|}{method} &
 \multicolumn{2}{>{\centering}m{0.07\textwidth}|}{sim\-u\-la\-tion} &
 tot. sys. &
 tot. \\
  & & & &
 \multicolumn{2}{>{\centering}m{0.07\textwidth}|}{(i)} &
 \multicolumn{2}{>{\centering}m{0.08\textwidth}|}{(ii)} &
 \multicolumn{2}{>{\centering}m{0.08\textwidth}|}{(iii)} &
 \multicolumn{2}{>{\centering}m{0.07\textwidth}|}{(iv)} &
 \multicolumn{2}{>{\centering}m{0.07\textwidth}|}{(v)} &
 & \\
\hline
$r^{04}_{00}$ & $ $ & $0.346$ & $0.018$ & $-$ & $0.001$ & $ $ & $0.000$ & $ $ & $0.001$ & $ $ & $0.003$ & $ $ & $0.007$ & $0.008$ & $ 0.022$ \\
$r^{1}_{1-1}$ & $-$ & $0.041$ & $0.023$ & $ $ & $0.033$ & $ $ & $0.008$ & $ $ & $0.007$ & $ $ & $0.016$ & $ $ & $0.000$ & $0.038$ & $ 0.045$ \\
Im $r^{2}_{1-1}$ & $ $ & $0.031$ & $0.023$ & $ $ & $0.029$ & $ $ & $0.015$ & $ $ & $0.020$ & $-$ & $0.031$ & $-$ & $0.002$ & $0.049$ & $ 0.054$ \\
\hline
Re $r^{5}_{10}$  & $ $ & $0.103$ & $0.008$ & $-$ & $0.004$ & $ $ & $0.003$ & $ $ & $0.005$ & $ $ & $0.006$ & $-$ & $0.001$ & $0.009$ & $ 0.012$ \\
Im $r^{6}_{10}$  & $-$ & $0.089$ & $0.007$ & $ $ & $0.004$ & $ $ & $0.007$ & $ $ & $0.000$ & $-$ & $0.012$ & $ $ & $0.001$ & $0.015$ & $ 0.017$ \\
Im $r^{7}_{10}$  & $ $ & $0.005$ & $0.081$ & $ $ & $0.115$ & $ $ & $0.024$ & $ $ & $0.000$ & $ $ & $0.009$ & $ $ & $0.002$ & $0.118$ & $ 0.143$ \\
Re $r^{8}_{10}$  & $ $ & $0.093$ & $0.072$ & $-$ & $0.019$ & $-$ & $0.004$ & $ $ & $0.000$ & $ $ & $0.016$ & $ $ & $0.001$ & $0.025$ & $ 0.076$ \\
\hline
Re $r^{04}_{10}$ & $ $ & $0.018$ & $0.011$ & $ $ & $0.001$ & $ $ & $0.010$ & $ $ & $0.004$ & $ $ & $0.008$ & $ $ & $0.004$ & $0.014$ & $ 0.018$ \\
Re $r^{1}_{10}$  & $-$ & $0.081$ & $0.016$ & $-$ & $0.003$ & $-$ & $0.019$ & $ $ & $0.000$ & $-$ & $0.009$ & $-$ & $0.002$ & $0.022$ & $ 0.027$ \\
Im $r^{2}_{10}$  & $ $ & $0.061$ & $0.015$ & $ $ & $0.000$ & $ $ & $0.010$ & $ $ & $0.006$ & $ $ & $0.017$ & $ $ & $0.004$ & $0.021$ & $ 0.026$ \\
$r^{5}_{00}$  & $ $ & $0.132$ & $0.014$ & $ $ & $0.005$ & $ $ & $0.009$ & $-$ & $0.012$ & $ $ & $0.035$ & $ $ & $0.010$ & $0.040$ & $ 0.043$ \\
$r^{1}_{00}$  & $-$ & $0.078$ & $0.028$ & $ $ & $0.010$ & $-$ & $0.031$ & $ $ & $0.016$ & $-$ & $0.011$ & $-$ & $0.008$ & $0.038$ & $ 0.048$ \\
Im $r^{3}_{10}$  & $ $ & $0.057$ & $0.052$ & $-$ & $0.048$ & $ $ & $0.019$ & $ $ & $0.000$ & $ $ & $0.052$ & $-$ & $0.001$ & $0.073$ & $ 0.090$ \\
$r^{8}_{00}$  & $ $ & $0.125$ & $0.130$ & $ $ & $0.125$ & $ $ & $0.077$ & $ $ & $0.000$ & $ $ & $0.016$ & $ $ & $0.001$ & $0.148$ & $ 0.197$ \\
\hline
$r^{5}_{11}$  & $-$ & $0.016$ & $0.009$ & $-$ & $0.002$ & $ $ & $0.028$ & $-$ & $0.009$ & $-$ & $0.013$ & $ $ & $0.006$ & $0.032$ & $ 0.034$ \\
$r^{5}_{1-1}$ & $-$ & $0.017$ & $0.012$ & $ $ & $0.020$ & $ $ & $0.003$ & $-$ & $0.008$ & $-$ & $0.011$ & $ $ & $0.002$ & $0.025$ & $ 0.027$ \\
Im $r^{6}_{1-1}$ & $ $ & $0.023$ & $0.011$ & $ $ & $0.010$ & $-$ & $0.006$ & $-$ & $0.005$ & $ $ & $0.012$ & $ $ & $0.001$ & $0.018$ & $ 0.021$ \\
Im $r^{7}_{1-1}$ & $ $ & $0.150$ & $0.111$ & $ $ & $0.147$ & $-$ & $0.012$ & $ $ & $0.000$ & $ $ & $0.079$ & $ $ & $0.000$ & $0.168$ & $ 0.201$ \\
$r^{8}_{11}$  & $-$ & $0.106$ & $0.078$ & $-$ & $0.030$ & $-$ & $0.001$ & $ $ & $0.000$ & $-$ & $0.047$ & $ $ & $0.002$ & $0.056$ & $ 0.096$ \\
$r^{8}_{1-1}$ & $-$ & $0.009$ & $0.101$ & $ $ & $0.108$ & $-$ & $0.054$ & $ $ & $0.000$ & $-$ & $0.028$ & $-$ & $0.002$ & $0.124$ & $ 0.160$ \\
\hline
$r^{04}_{1-1}$& $ $ & $0.022$ & $0.016$ & $ $ & $0.005$ & $ $ & $0.004$ & $-$ & $0.008$ & $ $ & $0.003$ & $ $ & $0.000$ & $0.010$ & $ 0.019$ \\
$r^{1}_{11}$  & $-$ & $0.025$ & $0.018$ & $-$ & $0.015$ & $-$ & $0.005$ & $ $ & $0.000$ & $ $ & $0.002$ & $ $ & $0.000$ & $0.016$ & $ 0.024$ \\
Im $r^{3}_{1-1}$ & $ $ & $0.095$ & $0.071$ & $ $ & $0.109$ & $ $ & $0.002$ & $ $ & $0.000$ & $ $ & $0.010$ & $ $ & $0.002$ & $0.110$ & $ 0.131$ \\
\hline
\end{tabular}
\end{center}
\end{table*}

\begin{table}[ht]
\renewcommand{\arraystretch}{1.2}
\begin{center}
\caption{\label{tab2} The measured 23 unpolarised and polarised $\omega$ SDMEs
 in bins of $Q^2$: $1.00 - 1.35 - 2.05 - 10.00$~(GeV/$c$)$^2$.
The first uncertainties are statistical, the second  systematic.}
\begin{tabular}{|c|r@{$\,\pm\,$}l@{$\,\pm\,$}l|r@{$\,\pm\,$}l@{$\,\pm\,$}l|r@{$\,\pm\,$}l@{$\,\pm\,$}l|}
\hline
 SDME &\multicolumn{3}{c|}{$\langle Q^{2}\rangle$ = 1.16 (GeV/$c$)$^2$}
         &\multicolumn{3}{c|}{$\langle Q^{2}\rangle$ = 1.64 (GeV/$c$)$^2$}
         &\multicolumn{3}{c|}{$\langle Q^{2}\rangle$ = 3.58 (GeV/$c$)$^2$} \\
\hline
$r^{04}_{00}$    &$  0.313$&$ 0.027$&$ 0.027 $&$  0.322$&$ 0.029$&$ 0.046 $&$  0.415$&$ 0.038$&$ 0.052$ \\
$r^1_{1-1}$      &$ -0.029$&$ 0.037$&$ 0.015 $&$ -0.011$&$ 0.037$&$ 0.036 $&$ -0.101$&$ 0.048$&$ 0.090$ \\
Im $r^2_{1-1}$   &$ -0.027$&$ 0.037$&$ 0.025 $&$  0.036$&$ 0.037$&$ 0.047 $&$  0.103$&$ 0.049$&$ 0.065$ \\
\hline
Re $r^5_{10}$    &$  0.120$&$ 0.012$&$ 0.008 $&$  0.086$&$ 0.013$&$ 0.013 $&$  0.101$&$ 0.017$&$ 0.050$ \\
Im $r^6_{10}$    &$ -0.102$&$ 0.012$&$ 0.009 $&$ -0.075$&$ 0.012$&$ 0.012 $&$ -0.088$&$ 0.016$&$ 0.033$ \\
Im $r^7_{10}$    &$  0.079$&$ 0.127$&$ 0.140 $&$  0.061$&$ 0.136$&$ 0.070 $&$ -0.150$&$ 0.167$&$ 0.182$ \\
Re $r^8_{10}$    &$  0.164$&$ 0.119$&$ 0.183 $&$ -0.131$&$ 0.112$&$ 0.116 $&$  0.340$&$ 0.153$&$ 0.144$ \\
\hline
Re $r^{04}_{10}$ &$  0.020$&$ 0.017$&$ 0.016 $&$  0.014$&$ 0.017$&$ 0.016 $&$  0.016$&$ 0.023$&$ 0.013$ \\
Re $r^1_{10}$    &$ -0.080$&$ 0.023$&$ 0.033 $&$ -0.074$&$ 0.025$&$ 0.055 $&$ -0.080$&$ 0.035$&$ 0.045$ \\
Im $r^2_{10}$    &$  0.064$&$ 0.025$&$ 0.012 $&$  0.066$&$ 0.025$&$ 0.018 $&$  0.069$&$ 0.032$&$ 0.046$ \\
$r^5_{00}$       &$  0.059$&$ 0.021$&$ 0.022 $&$  0.130$&$ 0.023$&$ 0.039 $&$  0.219$&$ 0.031$&$ 0.077$ \\
$r^1_{00}$       &$ -0.002$&$ 0.041$&$ 0.028 $&$ -0.090$&$ 0.043$&$ 0.073 $&$ -0.144$&$ 0.063$&$ 0.087$ \\
Im $r^3_{10}$    &$ -0.026$&$ 0.080$&$ 0.187 $&$  0.175$&$ 0.087$&$ 0.109 $&$ -0.038$&$ 0.109$&$ 0.062$ \\
$r^8_{00}$       &$ -0.064$&$ 0.213$&$ 0.404 $&$  0.178$&$ 0.209$&$ 0.166 $&$  0.224$&$ 0.245$&$ 0.101$ \\
\hline
$r^5_{11}$       &$  0.009$&$ 0.014$&$ 0.035 $&$ -0.028$&$ 0.015$&$ 0.041 $&$ -0.037$&$ 0.019$&$ 0.026$ \\
$r^5_{1-1}$      &$ -0.002$&$ 0.019$&$ 0.015 $&$ -0.028$&$ 0.020$&$ 0.052 $&$ -0.026$&$ 0.024$&$ 0.020$ \\
Im $r^6_{1-1}$   &$  0.021$&$ 0.018$&$ 0.016 $&$  0.038$&$ 0.019$&$ 0.024 $&$  0.008$&$ 0.024$&$ 0.017$ \\
Im $r^7_{1-1}$   &$  0.159$&$ 0.161$&$ 0.216 $&$ -0.006$&$ 0.186$&$ 0.079 $&$  0.330$&$ 0.246$&$ 0.150$ \\
$r^8_{11}$       &$  0.092$&$ 0.123$&$ 0.120 $&$ -0.260$&$ 0.132$&$ 0.126 $&$ -0.188$&$ 0.163$&$ 0.159$ \\
$r^8_{1-1}$      &$ -0.023$&$ 0.163$&$ 0.451 $&$ -0.186$&$ 0.169$&$ 0.191 $&$  0.231$&$ 0.203$&$ 0.300$ \\
\hline
$r^{04}_{1-1}$   &$  0.034$&$ 0.025$&$ 0.034 $&$  0.009$&$ 0.026$&$ 0.043 $&$  0.024$&$ 0.032$&$ 0.016$ \\
$r^1_{11}$       &$ -0.024$&$ 0.028$&$ 0.028 $&$ -0.013$&$ 0.029$&$ 0.046 $&$ -0.042$&$ 0.038$&$ 0.024$ \\
Im $r^3_{1-1}$   &$  0.167$&$ 0.106$&$ 0.229 $&$  0.082$&$ 0.119$&$ 0.162 $&$  0.024$&$ 0.161$&$ 0.131$ \\
\hline
\end{tabular}
\end{center}
\end{table}

\begin{table*}[hbt!]
\renewcommand{\arraystretch}{1.2}
\begin{center}
\caption{\label{tab3} The measured 23 unpolarised and polarised $\omega$ SDMEs
 in bins of $p^{2}_{\rm T} $: $0.01 - 0.07 - 0.19 - 0.50$ (GeV/$c$)$^2$.
The first uncertainties are statistical, the second  systematic.}
\begin{tabular}{|c|r@{$\,\pm\,$}l@{$\,\pm\,$}l|r@{$\,\pm\,$}l@{$\,\pm\,$}l|r@{$\,\pm\,$}l@{$\,\pm\,$}l|}
\hline
 SDME &\multicolumn{3}{c|}{$\langle p^{2}_{\rm T}\rangle$ = 0.037 (GeV/$c$)$^2$}
         &\multicolumn{3}{c|}{$\langle p^{2}_{\rm T}\rangle$ = 0.124 (GeV/$c$)$^2$}
         &\multicolumn{3}{c|}{$\langle p^{2}_{\rm T}\rangle$ = 0.31  (GeV/$c$)$^2$}\\
\hline
$r^{04}_{00}$   &$ 0.272$&$ 0.027$&$ 0.016$&$ 0.310$&$ 0.033$&$ 0.025$&$ 0.479$&$ 0.035$&$ 0.045$ \\
$r^1_{1-1}$     &$-0.097$&$ 0.037$&$ 0.016$&$-0.058$&$ 0.040$&$ 0.034$&$ 0.051$&$ 0.043$&$ 0.067$ \\
Im $r^2_{1-1}$  &$ 0.095$&$ 0.037$&$ 0.020$&$ 0.017$&$ 0.042$&$ 0.046$&$-0.032$&$ 0.042$&$ 0.033$ \\
\hline
Re $r^5_{10}$   &$ 0.077$&$ 0.013$&$ 0.004$&$ 0.113$&$ 0.014$&$ 0.022$&$ 0.123$&$ 0.016$&$ 0.050$ \\
Im $r^6_{10}$   &$-0.090$&$ 0.011$&$ 0.012$&$-0.080$&$ 0.013$&$ 0.006$&$-0.102$&$ 0.015$&$ 0.029$ \\
Im $r^7_{10}$   &$-0.044$&$ 0.123$&$ 0.024$&$ 0.046$&$ 0.145$&$ 0.385$&$-0.009$&$ 0.159$&$ 0.023$ \\
Re $r^8_{10}$   &$ 0.120$&$ 0.122$&$ 0.052$&$ 0.168$&$ 0.120$&$ 0.136$&$-0.016$&$ 0.142$&$ 0.082$ \\
\hline
Re $r^{04}_{10}$&$ 0.001$&$ 0.017$&$ 0.016$&$ 0.034$&$ 0.019$&$ 0.021$&$ 0.033$&$ 0.022$&$ 0.033$ \\
Re $r^1_{10}$   &$-0.044$&$ 0.025$&$ 0.030$&$-0.068$&$ 0.027$&$ 0.009$&$-0.136$&$ 0.031$&$ 0.089$ \\
Im $r^2_{10}$   &$ 0.037$&$ 0.023$&$ 0.043$&$ 0.036$&$ 0.027$&$ 0.037$&$ 0.116$&$ 0.030$&$ 0.062$ \\
$r^5_{00}$      &$ 0.044$&$ 0.022$&$ 0.024$&$ 0.151$&$ 0.025$&$ 0.039$&$ 0.217$&$ 0.029$&$ 0.051$ \\
$r^1_{00}$      &$-0.089$&$ 0.040$&$ 0.026$&$-0.038$&$ 0.051$&$ 0.042$&$-0.080$&$ 0.058$&$ 0.056$ \\
Im $r^3_{10}$   &$ 0.048$&$ 0.081$&$ 0.167$&$-0.009$&$ 0.094$&$ 0.089$&$ 0.089$&$ 0.106$&$ 0.060$ \\
$r^8_{00}$      &$-0.035$&$ 0.185$&$ 0.094$&$ 0.266$&$ 0.230$&$ 0.261$&$ 0.169$&$ 0.270$&$ 0.208$ \\
\hline
$r^5_{11}$      &$-0.008$&$ 0.015$&$ 0.025$&$-0.028$&$ 0.016$&$ 0.014$&$-0.014$&$ 0.017$&$ 0.034$ \\
$r^5_{1-1}$     &$-0.002$&$ 0.019$&$ 0.013$&$-0.031$&$ 0.021$&$ 0.024$&$-0.012$&$ 0.022$&$ 0.021$ \\
Im $r^6_{1-1}$  &$ 0.009$&$ 0.018$&$ 0.030$&$ 0.054$&$ 0.021$&$ 0.016$&$ 0.006$&$ 0.021$&$ 0.008$ \\
Im $r^7_{1-1}$  &$ 0.289$&$ 0.171$&$ 0.125$&$-0.098$&$ 0.202$&$ 0.312$&$ 0.159$&$ 0.221$&$ 0.144$ \\
$r^8_{11}$      &$ 0.079$&$ 0.122$&$ 0.115$&$-0.260$&$ 0.144$&$ 0.222$&$-0.222$&$ 0.154$&$ 0.089$ \\
$r^8_{1-1}$     &$-0.005$&$ 0.161$&$ 0.094$&$-0.268$&$ 0.183$&$ 0.134$&$ 0.280$&$ 0.190$&$ 0.204$ \\
\hline
$r^{04}_{1-1}$  &$ 0.002$&$ 0.025$&$ 0.024$&$ 0.058$&$ 0.029$&$ 0.017$&$ 0.009$&$ 0.030$&$ 0.024$ \\
$r^1_{11}$      &$-0.023$&$ 0.029$&$ 0.057$&$-0.037$&$ 0.032$&$ 0.023$&$-0.021$&$ 0.034$&$ 0.024$ \\
Im $r^3_{1-1}$  &$ 0.097$&$ 0.107$&$ 0.278$&$-0.069$&$ 0.131$&$ 0.171$&$ 0.263$&$ 0.145$&$ 0.075$ \\
\hline
\end{tabular}
\end{center}
\end{table*}

\begin{table*}[hbt!]
\renewcommand{\arraystretch}{1.2}
\begin{center}
\caption{\label{tab4} The measured 23 unpolarised and polarised $\omega$ SDMEs
 in bins of $W$: $5.00 - 6.4 - 7.9 - 17.0$~GeV/$\it {c}^2$.
The first uncertainties are statistical, the second  systematic.}
\begin{tabular}{|c|r@{$\,\pm\,$}l@{$\,\pm\,$}l|r@{$\,\pm\,$}l@{$\,\pm\,$}l|r@{$\,\pm\,$}l@{$\,\pm\,$}l|}
\hline
 SDME &\multicolumn{3}{c|}{$\langle W\rangle$ = 5.87 GeV/$c^2$}
         &\multicolumn{3}{c|}{$\langle W\rangle$ = 7.06 GeV/$c^2$}
         &\multicolumn{3}{c|}{$\langle W\rangle$ = 9.90 GeV/$c^2$}\\
\hline
$r^{04}_{00}$    &$  0.315$&$ 0.034$&$ 0.012 $&$  0.355$&$ 0.029$&$ 0.055 $&$  0.333$&$ 0.027$&$ 0.035$ \\
$r^1_{1-1}$      &$ -0.134$&$ 0.043$&$ 0.003 $&$ -0.044$&$ 0.036$&$ 0.033 $&$  0.052$&$ 0.038$&$ 0.047$ \\
Im $r^2_{1-1}$   &$  0.139$&$ 0.044$&$ 0.046 $&$  0.037$&$ 0.036$&$ 0.024 $&$ -0.099$&$ 0.038$&$ 0.033$ \\
\hline
Re $r^5_{10}$    &$  0.066$&$ 0.015$&$ 0.021 $&$  0.102$&$ 0.012$&$ 0.010 $&$  0.125$&$ 0.012$&$ 0.016$ \\
Im $r^6_{10}$    &$ -0.071$&$ 0.014$&$ 0.003 $&$ -0.072$&$ 0.011$&$ 0.031 $&$ -0.115$&$ 0.012$&$ 0.015$ \\
Im $r^7_{10}$    &$  0.270$&$ 0.287$&$ 0.122 $&$  0.183$&$ 0.164$&$ 0.184 $&$ -0.099$&$ 0.082$&$ 0.065$ \\
Re $r^8_{10}$    &$  0.477$&$ 0.280$&$ 0.056 $&$  0.207$&$ 0.155$&$ 0.215 $&$  0.046$&$ 0.072$&$ 0.029$ \\
\hline
Re $r^{04}_{10}$ &$  0.039$&$ 0.021$&$ 0.010 $&$  0.013$&$ 0.017$&$ 0.022 $&$  0.007$&$ 0.017$&$ 0.016$ \\
Re $r^1_{10}$    &$ -0.092$&$ 0.030$&$ 0.034 $&$ -0.100$&$ 0.024$&$ 0.040 $&$ -0.051$&$ 0.026$&$ 0.018$ \\
Im $r^2_{10}$    &$  0.073$&$ 0.029$&$ 0.038 $&$  0.056$&$ 0.025$&$ 0.017 $&$  0.046$&$ 0.025$&$ 0.021$ \\
$r^5_{00}$       &$  0.154$&$ 0.026$&$ 0.023 $&$  0.145$&$ 0.023$&$ 0.019 $&$  0.081$&$ 0.023$&$ 0.007$ \\
$r^1_{00}$       &$ -0.114$&$ 0.053$&$ 0.069 $&$ -0.003$&$ 0.043$&$ 0.040 $&$ -0.104$&$ 0.043$&$ 0.018$ \\
Im $r^3_{10}$    &$  0.162$&$ 0.199$&$ 0.124 $&$  0.295$&$ 0.112$&$ 0.075 $&$ -0.001$&$ 0.054$&$ 0.082$ \\
$r^8_{00}$       &$ -0.834$&$ 0.467$&$ 0.047 $&$  0.451$&$ 0.282$&$ 0.138 $&$  0.044$&$ 0.129$&$ 0.133$ \\
\hline
$r^5_{11}$       &$  0.023$&$ 0.017$&$ 0.038 $&$ -0.027$&$ 0.015$&$ 0.024 $&$ -0.040$&$ 0.014$&$ 0.021$ \\
$r^5_{1-1}$      &$ -0.025$&$ 0.022$&$ 0.011 $&$ -0.024$&$ 0.019$&$ 0.046 $&$ -0.006$&$ 0.019$&$ 0.005$ \\
Im $r^6_{1-1}$   &$  0.030$&$ 0.021$&$ 0.007 $&$  0.023$&$ 0.018$&$ 0.020 $&$  0.026$&$ 0.018$&$ 0.018$ \\
Im $r^7_{1-1}$   &$ -0.105$&$ 0.435$&$ 0.235 $&$ -0.173$&$ 0.236$&$ 0.082 $&$  0.210$&$ 0.112$&$ 0.135$ \\
$r^8_{11}$       &$  0.264$&$ 0.320$&$ 0.043 $&$ -0.177$&$ 0.181$&$ 0.192 $&$ -0.127$&$ 0.082$&$ 0.060$ \\
$r^8_{1-1}$      &$  0.124$&$ 0.414$&$ 0.193 $&$ -0.177$&$ 0.222$&$ 0.155 $&$ -0.004$&$ 0.108$&$ 0.152$ \\
\hline
$r^{04}_{1-1}$   &$  0.027$&$ 0.031$&$ 0.015 $&$  0.022$&$ 0.025$&$ 0.028 $&$  0.025$&$ 0.024$&$ 0.018$ \\
$r^1_{11}$       &$ -0.023$&$ 0.034$&$ 0.005 $&$ -0.060$&$ 0.028$&$ 0.011 $&$  0.006$&$ 0.028$&$ 0.005$ \\
Im $r^3_{1-1}$   &$  0.406$&$ 0.297$&$ 0.100 $&$  0.015$&$ 0.168$&$ 0.042 $&$  0.088$&$ 0.072$&$ 0.085$ \\

\hline
\end{tabular}
\end{center}
\end{table*}


\clearpage


\begin{thebibliography}{50}
\bibitem{gpd1} D.~M\"uller et al., Fortschr. Phys.  {\bf 42}, 101~(1994).
\bibitem{gpd2} X.~Ji, Phys. Rev. Lett. {\bf 78}, 610~(1997).
\bibitem{gpd3} X.~Ji, Phys. Rev. D {\bf 55}, 7114~(1997).
\bibitem{gpd4} A.V.~Radyushkin, Phys. Lett. B {\bf 385}, 333 (1996).
\bibitem{Radyushkin:1996ru} A.V.~Radyushkin, Phys. Rev. D {\bf 56}, 5524~(1997).
\bibitem{Collins:1996fb} J.C.~Collins, L. Frankfurt, M. Strikman, Phys. Rev. D {\bf 56}, 2982~(1997).
\bibitem{Martin-1997} A.D.~Martin, M.G.~Ryskin, T.~Teubner, Phys. Rev. {\bf 55}, 4329~(1997).
\bibitem{Goloskokov:2005} S.V. Goloskokov, P. Kroll, Eur. Phys. J. C {\bf 42}, 281 (2005).
\bibitem{Goloskokov:2008} S.V. Goloskokov, P. Kroll, Eur. Phys. J. C {\bf 53}, 367 (2008).
\bibitem{Goloskokov:2009} S.V. Goloskokov, P. Kroll, Eur. Phys. J. C {\bf 59}, 809 (2009).
\bibitem{GK:epjC-2014} S.V.~Goloskokov, P.~Kroll, Eur. Phys. J. C {\bf 74}, 2725 (2014).
\bibitem{GK:epjA-2014} S.V.~Goloskokov, P.~Kroll, Eur. Phys. J. A {\bf 50}, 146 (2014).
\bibitem{HERMES:2014} A.~Airapetian et al., (HERMES Collaboration), Eur. Phys. J. C {\bf 74}, 3110 (2014); Erratum: Eur. Phys. J. C {\bf 76}, 162 (2016).
\bibitem{DC-24} A.~Airapetian et al., (HERMES Collaboration), Eur. Phys. J. C {\bf 62}, 659~(2009).
%\bibitem{Bauer} T.H.~Bauer, R.D.~Spital, D.R.~Yennie, F.M.~Pipkin, Rev. Mod. Phys. {\bf 50}, 261 (1978).
\bibitem{joos} P.~Joos et al., Nucl. Phys. B {\bf 122}, 365 (1977).
\bibitem{clas}  L.~Morand et al., (CLAS Collaboration), Eur. Phys. J. A {\bf 24}, 445~(2005).
%\bibitem{zeus} J.~Breitweg et al., (ZEUS Collaboration), Phys. Lett. B {\bf 487}, 273 (2000).
\bibitem{COMPASS-asy} C.~Adolph et al., (COMPASS Collaboration), Nucl. Phys. B {\bf 915}, 454 (2017).
\bibitem{Schill} K.~Schilling and G.~Wolf, Nucl. Phys. B {\bf  61}, 381(1973).
\bibitem{Diehl} M.~Diehl, JHEP  {\bf 0709}, 064 (2007). 
%\bibitem{joos} P.~Joos et al., Nucl. Phys.  B {\bf 122}, 365 (1977).
\bibitem{comp1} P. Abbon et al., (COMPASS Collaboration), Nucl. Instrum. Meth. A {\bf 557}, 455 (2007).
\bibitem{comp2} P. Abbon et al., (COMPASS Collaboration), Nucl. Instrum. Meth. A {\bf 779}, 69 (2015).
\bibitem{comp3} F. Gautheron et al., (COMPASS Collaboration), SPSC-P-340, CERN-SPSC-2019-014.
\bibitem{psz} P.~Sznajder, PhD thesis, National Centre For Nuclear Research, Otwock – \'{S}wierk, March 2015. 
\bibitem{Comptune} C.~Adolph et al., (COMPASS Collaboration), Phys. Lett. B {\bf 718}, 922 (2013).
\bibitem{hepg3} T. Szameitat, PhD thesis, University of Freiburg (2017), doi:10.6094/UNIFR/11686.
\bibitem{hepg1} A. Sandacz and P. Sznajder, ``HEPGEN - generator for hard exclusive leptoproduction'', (2012), arXiv:1207.0333.
\bibitem{hepg2} C. Regali, PhD thesis, University of Freiburg (2016), doi:10.6094/UNIFR/11449.
\bibitem{comnote} E. Burtin, N. d'Hose, O.A. Grajek and A. Sandacz, ``Angular distributions and $R = \sigma_L / \sigma_T$ for exclusive $\rho ^0$ production'', private communication.
\bibitem{DVCSpaper} R.~Akhunzyanov et al., (COMPASS Collaboration), Phys. Lett. B {\bf 793}, 188 (2019).
\bibitem{pi0paper} M.G.~Alexeev et al., (COMPASS Collaboration), Phys. Lett. B {\bf 805}, 135454 (2020). 
%arXiv:1903.12030.  
%\bibitem{Fran} L. Frankfurt et al., Phys. Rev. D {\bf 54}, 3194 (1996)
%\bibitem{lastGK} S.V.~Goloskokov, private communication.
%\bibitem{sigmat} S. Donnachie, G. Dosch, P. Landshoff and O. Nachtmann,{ \it
%Pomeron Physics and QCD (Cambride University press, New York, 2002).}
\end{thebibliography}
\end{document}